# Chemical pressure effects on structural, dielectric and magnetic properties of solid solutions $Mn_{3-x}Co_xTeO_6$

S. A. Ivanov[ab], R. Mathieu[b*], P. Nordblad[b], C. Ritter[c], R. Tellgren[d], N. Golubko[a], A. Mosunov[a], E. D. Politova[a], M. Weil[e]

a- Center of Materials Science, Karpov Institute of Physical Chemistry, Moscow, 105064, Russia
b- Department of Engineering Sciences, Uppsala University, Box 534, SE-751 21 Uppsala, Sweden
c- Institute Laue-Langevin, Grenoble, France
d- Department of Chemistry, Uppsala University, Box 538, SE-751 21 Uppsala, Sweden
e- Institute for Chemical Technologies and Analytics, Vienna University of Technology, A-1060 Vienna, Austria

# Abstract

The effects of $Co^{2+}$ doping on the structural, magnetic and dielectric properties of the multiferroic frustrated antiferromagnet $Mn_3TeO_6$ have been investigated. Ceramic samples of the solid solution series $Mn_{3-x}Co_xTeO_6$ were prepared by a solid-state reaction route. X-ray and neutron powder diffraction and electron microscopy techniques were combined with calorimetric, dielectric and magnetic measurements to investigate the dependence of the crystal structure and physical properties on temperature and composition. It is shown that the compounds with $x \leq 2.4$ adopt the trigonal corundum-related structure of pure $Mn_3TeO_6$ (space group $R\bar{3}$) in the temperature range 5-295 K and that the lattice parameters $a$ and $c$ and the unit-cell volume $V$ decrease linearly with increasing $Co^{2+}$ concentration. The low-temperature magnetic susceptibility and heat capacity data evidence the antiferromagnetic ordering of all samples. The Neel temperature linearly increases with $Co^{2+}$ concentration $x$. Curie-Weiss fits of the high temperature susceptibility indicate that the magnetic frustration decreases with $x$. The derived magnetic structure of $Mn_3TeO_6$ can be described as an incommensurately modulated magnetic spin state with k = [0, 0, $k_z$] and an elliptical spin-spiral order of spins within the chains of $MnO_6$ octahedra. With increasing $Co^{2+}$ concentration the propagation vector $k_z$ changes from 0.453 ($x$ = 0) to 0.516 ($x$ = 2.4). The magnetic anisotropy changes as well, leading to a reorientation of the spiral-basal plane. A possible coexistence of long-range order of electrical dipoles and magnetic moments in $Mn_{3-x}Co_xTeO_6$ is discussed.

* Corresponding author.
*E-mail address:* roland.mathieu@angstrom.uu.se

## 1. Introduction

Recent intense research on multiferroic materials has led to the discovery of new types of compounds with spin and dipole ordering and to a better understanding of the fundamental physical processes and interactions behind their complex physical behavior [1-4]. Many of these single-phase materials possess long-wavelength geometrically frustrated spin networks. In this class of antiferromagnetic (AFM) materials, incommensurate (ICM) magnetic structures with spiral-spin order occur in which any inversion-symmetry breaking event can induce a spontaneous electric polarization via the spin-orbit coupling. However, the origin of the magnetoelectric coupling in these systems is complicated and essential aspects of their properties still are not fully understood. Also, only few single-phase multiferroics have so far been discovered [5-17]. Therefore, it is of great interest to find new single-phase materials that show multiferroic properties. Up to date the search for new multiferroic materials has mainly been focused on perovskite-based materials due to their compositional flexibility and because the mechanisms that govern the properties of this structure type are well understood [15-17].

However, in recent years also transition metal orthotellurates, $A_3TeO_6$, with corundum-related structures, where $A$ = Mn, Co, Ni, Cu, have been extensively studied as possible multiferroics [18-32]. This family of materials is known to exhibit various properties as a result of compositional flexibility. The structure of many $A_3TeO_6$ ($A_2A$TeO$_6$) phases is related to the double perovskites $A_2BB'O_6$ [33]; however, the presence of additional crystallographically non-equivalent sites for the magnetic cations provides extra degrees of freedom for manipulation with the structure. Among these potential spin-spiral multiferroics, Mn$_3$TeO$_6$ (MTO), in which non-Jahn-Teller Mn$^{2+}$ ($3d^5$) cations carry S=5/2 spins, is a promising example. MTO was first described and prepared in powder form by Bayer [34] who found the compound to be isotypic to the corundum-related Mg$_3$TeO$_6$ family. The interest in $A_3TeO_6$ phases also increased after Kosse *et al.* [35,36] had identified them as a new class of ferroelectrics.

It was recently reported [27] that MTO enters a complex long-range magnetically ordered state below 23 K. In this structure, the incommensurate magnetic propagation vector $k$ = [0, 0, 0.4302(1)] splits the unique Mn site into two magnetically different orbits. One orbit forms a perfect helix with the spiral axis along the $c$-axis while the other orbit has a sine wave character along the $c$-axis. The loss of inversion symmetry due to the helical spin ordering may introduce ferroelectric polarization in MTO.

Co$_3$TeO$_6$ (CTO) belongs to the family of transition-metal orthotellurates. It crystallizes in the monoclinic space group $C2/c$ [21,28]. The Co$^{2+}$ ions are located in five crystallographically distinct sites, four of which are octahedrally coordinated and one tetrahedrally coordinated; the corresponding CoO$_6$ and CoO$_4$ polyhedra are connected by corner-, edge-, and face-sharing. through Co-O bonds. The exchange interactions between the Co$^{2+}$ ions in Co$_3$TeO$_6$ are sufficiently strong to result in long-range magnetic ordering. This compound exhibits magnetic field driven electric polarization at low temperatures, indicating strong coupling between magnetic and electric dipoles [37].

In this study $A$-site substitution of Mn by Co in the solid solution series Mn$_{3-x}$Co$_x$TeO$_6$ ($0 \leq x \leq 3$) (MCTO) is investigated with regard to structural, dielectric and magnetic properties. We surprisingly find that all samples up to $x$=2.4 adopt the trigonal corundum-related structure of pure Mn$_3$TeO$_6$ (space group $R\bar{3}$), i.e. not even substitution of 80 % of the Mn$^{2+}$ ions with Co$^{2+}$ transfers MCTO to the monoclinic structure of pure Co$_3$TeO$_6$. The lattice parameters of MCTO decrease linearly, the magnetic ordering temperature increases and the level of magnetic frustration decreases with increasing Co$^{2+}$ content.

## 2. Experimental

### 2.1 Sample preparation
#### 2.1.1 Single crystals
Single crystals of MTO were grown by chemical transport reactions [23]. A mixture of MnO and $TeO_3$ in the stoichiometric ratio 3:1 was thoroughly ground, pressed into a pellet and placed in a silica ampoule which was evacuated, sealed, and heated within 3 h to 1103 K and kept at this temperature for 3 days. This material was mixed with $PtCl_2$ and loaded in an evacuated and sealed silica ampoule which was heated in a temperature gradient 1103→1023 K. At this temperature, $PtCl_2$ decomposes with release of $Cl_2$ which then serves as the actual transport agent. After 5 days, a few amber coloured crystals of MTO with a plate-like shape and an edge-length up to 0.8 mm had formed in the colder part of the ampoule.

Single crystals of CTO were grown by chemical transport reactions [28] from CoO and $TeO_3$ as starting materials. CoO was prepared by heating $Co(NO_3)_2 \cdot 6H_2O$ in a platinum crucible at 1473 K for one hour and subsequent quenching in an ice bath; $TeO_3$ was prepared by heating $H_6TeO_6$ at 673 K for 12 hours. A mixture of CoO and $TeO_3$ in the stoichiometric ratio 3:1 was thoroughly ground and, together with $PtCl_2$, placed in a silica ampoule which was evacuated, sealed, and heated in a temperature gradient 1098 →1028 K. After 5 days the transport reaction was completed and dark blue to black crystals of $Co_3TeO_6$, mostly with a prismatic pinacoidal form and edge-lengths up to 5 mm, had formed in the colder part of the ampoule.

#### 2.1.2 Ceramic samples
High quality ceramic samples of $Mn_{3-x}Co_xTeO_6$ ($0 \leq x \leq 2.7$) were prepared by a solid state reaction route following a method described elsewhere [27, 32]. High purity $Mn_2O_3$, $Co(NO_3)_2 \cdot 6H_2O$ and $H_6TeO_6$ were used as starting materials. The raw materials were weighed in appropriate proportions for the nominal $Mn_{3-x}Co_xTeO_6$ composition. The homogenized stoichiometric mixtures were calcined at 770 K for 7 hours, ground in an agate mortar, pressed into pellets, placed into corundum crucibles, heated with a temperature interval of 100 K up to 1200 K ($x = 0$), 1000 K ($0 < x < 3$) and 970 K ($x = 3$) in air in a muffle furnace, held at this temperature for 8 hours and slowly cooled to room temperature. Under these conditions of synthesis only negligible oxygen deficiency is expected.

### 2.2. Chemical composition
The chemical composition of the prepared crystals and ceramic samples was analyzed by energy-dispersive X-ray spectroscopy (EDS) using a JEOL 840A scanning electron microscope and the INCA 4.07 (Oxford Instruments) software testing up to 20 different spots. As further confirmed by X-ray and neutron powder diffraction measurements, the refined concentration ratios of Mn, Co and Te of all samples were found to be close to the target compositions within the instrumental resolution (0.05-0.1 $\%_{wt}$). The determined concentrations of cations are reasonably consistent with the expected (theoretical) values; the discrepancy is generally less than 1 $\%_{wt.}$ for Te and < 1.5 $\%_{wt.}$ for Mn and Co. The analytical total is remarkably close to 100%. This suggests that our quantification procedure is appropriate for the measurements of cation contents of MCTO samples with the accuracy of ~1-1.5 $\%_{wt.}$. In the following sections, we will use the nominal composition to distinguish between different substitution levels.

In order to analyze the decomposition reaction and possible phase transformations, thermogravimetric and differential thermal analysis (TGA-DTA) were conducted in a Setaram Labsys TGA-DTA/DSC analyzer, from 300 to 1300 K in flowing air with a constant heating and cooling rate of 5-10 K/min. One of two symmetrical alumina crucibles was filled with the sample powder (around 50 mg); the other crucible contained alumina powder, used as a reference. The temperature was monitored using a thermocouple with accuracy of ± 0.1 K.

## 2.3. X-ray diffraction

Structure analysis of MTO and CTO single crystals was performed on a SMART Bruker diffractometer and the results were published earlier [27,28]. The phase identification and purity of MCTO powder samples were checked from X-ray powder diffraction (XRPD) patterns obtained with a D-5000 diffractometer using CuK$_\alpha$ radiation. Ceramic samples of MCTO were crushed into powders in an agate mortar and suspended in ethanol. A silicon wafer was covered with several drops of the resulting suspension, leaving randomly oriented crystallites after drying. The XRPD data for Rietveld analysis were collected at room temperature on a Bruker D8 Advance diffractometer (Vantec position-sensitive detector, Ge-monochromatized CuK$_\alpha$ radiation, Bragg-Brentano geometry, DIFFRACT plus software) in the 2θ range of 10-152° with a step size of 0.02° (counting time was 15 s per step). The slit system was selected to ensure that the X-ray beam was completely within the sample for all 2θ angles. The diffractometer was calibrated with Si SRM640b standard as a reference material.

## 2.4. Second harmonic generation (SHG) measurements.

The material was characterized by SHG measurements in reflection geometry, using a pulsed Nd:YAG laser (λ=1.064 μm). The SHG signal $I2\omega$ was measured from the polycrystalline samples relative to an α-quartz standard at room temperature in the Q-switching mode with a repetition rate of 4 Hz.

## 2.5. Dielectric measurements

Dielectric properties of single crystals of MTO and CTO were studied by dielectric spectroscopy in the frequency range of 1 kHz- 1MHz and the temperature range 300-1000 K (impedance bridge Agilent 14284 A, 2 K/min).
Powders of MCTO were used to make dense ceramics suitable for dielectric permittivity measurements. These disk-shaped samples were sintered at temperatures in the range of 900-1200 K for 8 h in air. The relative density of all sintered samples was around 80% of the theoretical density. Prior to these measurements, silver electrodes were attached on the circular faces of the disks using silver paste. The measurements on these samples were conducted in the frequency range 1-10$^4$ kHz and in the temperature range of 295-1000 K at heating and cooling rates of 1 K/min.

## 2.6. Magnetic and heat capacity measurements

The magnetization experiments were performed in a Quantum Design MPMSXL 5 T SQUID magnetometer. The magnetization (M) of single-crystal and ceramic samples was recorded as a function of temperature T in a small magnetic field (20 or 50 Oe) using zero-field-cooled (ZFC) and field-cooled (FC) protocols. Additional FC experiments in larger magnetic fields (5000 Oe) were performed for some of the samples. Specific heat (C) measurements were performed using a relaxation method between 2 K and 100 K on a Physical Property Measurement System (PPMS6000).

## 2.7. Neutron powder diffraction

Because the neutron scattering lengths of Mn, Co and Te are different (b(Mn)=-3.73 fm, b(Co)=2.49 fm, b(Te)=5.80 fm), the chemical composition of MCTO can be determined by neutron powder diffraction (NPD) with good precision. The neutron scattering length of oxygen anions (b(O)=5.803 fm) is comparable to those of the cations and NPD provides accurate information on their position and composition. The neutron diffraction experiments on MCTO samples were performed at the Institute Laue-Langevin (Grenoble, France) on the powder diffractometer D1A (wavelength 1.91 Å) in the 2θ-range 10– 156.9° with a step size of 0.1°. The powdered sample was inserted in a cylindrical vanadium container. A helium cryostat was used to collect data in the temperature range 5-295 K. Nuclear and magnetic refinements were performed by the Rietveld method using the FULLPROF software [38]. The diffraction peaks were described by a pseudo-Voigt profile function, with a Lorentzian contribution to the Gaussian peak shape. A peak asymmetry correction was made for angles below 35° (2θ). Background intensities were estimated by interpolating between up to 40 selected points (low temperature NPD experimental data) or described by a polynomial with six coefficients. The IVTON software [39] was employed to characterize the coordination spheres of the cations, the volumes of the coordination polyhedra and the displacements of the cations from the centres of the octahedra. The magnetic structure was refined in space group $R1$ as an independent phase in which only $Mn^{2+}$ and $Co^{2+}$ cations were included. Several magnetic models were tried in the refinement, each employing one additional refinement parameter, corresponding to the magnitude of the magnetic moment. Each structural model was refined to convergence. The model for which the structural refinement was stable and the $R$ factors were minimal was chosen as the final model. The K-search software was used for determination of a propagation vector [38]. We used the program BASIREPS [40], which calculates the existence of the irreducible representations for our special case. This program also extracts command lines for the magnetic refinement in the frame of the FULLPROF software [38].

## 3. Results

### 3.1 Structural Characterization

From thermal analyses it was found that formation of MCTO phases starts in the temperature range of 900-920 K, and single-phase samples are obtained after annealing at 1000 K. According to the elemental analyses performed on 20 different crystallites, the metal composition of the MCTO samples is quite close to the expected ratios and permits to conclude that the sample stoichiometry is the nominal one. The oxygen content, as determined by thermogravimetric analysis, is also in agreement with the proposed composition. The microstructure of the obtained powders, observed by scanning electron microscopy, revealed uniform and fine grain distribution.

The initial crystallographic characterization of the MCTO samples was performed by XRPD analysis of room temperature diffractograms. These have very narrow diffraction peaks without splitting or extra reflections (Fig.1a) which show that the prepared samples are single phase. It was found that all members of the series with $x < 2.4$ crystallize with trigonal symmetry and the structures could be successfully refined by the Rietveld method on the basis of the $R\bar{3}$ structural model, proposed for MTO in [23], using a random distribution of the Mn and Co ions on the Wyckoff (18 $f$) position. With this model, all structures of the prepared MCTO samples could be successfully refined (see Fig.1). All atomic positions are fully occupied. No structural phase transition occurs at room temperature with increasing Co content. For $x > 2.4$ ($x = 2.7$) a mixture of the trigonal ($R\bar{3}$) and monoclinic ($C2/c$) phases coexist at room temperature, in ratios depending of synthesis conditions.

Figure 2 shows the refined lattice parameters and the calculated cell volume of MCTO as a function of the nominal molar concentration $x$. The regular decreasing trend in the lattice parameters and unit cell volume is attributed to the replacement of $Mn^{2+}$ (0.83 Å) ions by the smaller $Co^{2+}$ (0.745 Å) ions [41], in the host system. Note that we have included data for $x$ = 2.7 in the figure for comparison. Moreover, Co doping increases the diffraction intensity in the X-ray diffraction patterns. The $A$-O ($A$ = Mn, Co) and Te-O bond lengths calculated from the refined lattice parameters and atomic coordinates are in good agreement with those observed in the literature [23]. Possible changes in the oxidation states of $Mn^{2+}$ and $Co^{2+}$ due to the substitution should be reflected in the transition-metal oxygen bond lengths along this series. Only minor changes in the Mn-O bond length with variation of the Co-content are observed. The values of the bond valence calculation [42] confirm the observed trend of the bond lengths. Furthermore, the corresponding bond valence sum calculations are consistent with the presence of $Mn^{2+}$, $Co^{2+}$, $Te^{6+}$ and $O^{2-}$ ions; any partial transition to a higher oxidation state for the $Mn^{2+}$ and $Co^{2+}$ cations could not be resolved.

Due to the occupation of the $A$ site by varying mixtures of $Mn^{2+}$ and $Co^{2+}$ cations, a significant variation of the unit cell parameters and the distortion indices of the coordination polyhedra is observed (Table 1). As expected from the ionic radii, the unit cell dimensions and unit cell volumes of MCTO decrease with increasing Co doping. The average <$A$–O> bond length becomes shorter with increasing Co concentration, whereas the average <Te-O> bond length remains unchanged within the accuracy of determination (Table 2).

The absence of compositionally driven changes in the <Te–O> bond length for the MCTO compounds indicates that $Co^{2+}$ ions do not replace the smaller Te cations.

The structure of MCTO consists of distorted coordination polyhedra, similar to those occurring in MTO.

The $AO_6$ octahedra have relatively large bond length and bond angle distortion parameters which increase regularly with Co doping. The $TeO_6$ octahedra are not distorted.

The $A$ ions have a greater displacement from the centres of the coordination polyhedra than the Te atoms, however, both these shifts decrease with Co doping.

In the case of MCTO the deviation from cubic symmetry is very small, the rhombohedral parameter $a_R$ and angle $\alpha_R$ are 6.24 Å and 90.67° for $Mn_3TeO_6$ decreasing with Co concentration. It can be seen from Table 1 that the lattice ratio $c_H/a_H$ for a large doping content is very close to the ratio $\sqrt{3} : \sqrt{2} = 1.225$, *viz.* as for a cube in the hexagonal setting. Such small deviations from cubic symmetry are usually characteristic for compounds, which may show a displacive transformation to a cubic high-temperature form (c.f. the cubic structure of $Cu_3TeO_6$). Experiments on MCTO samples using a high temperature XRPD diffractometer showed that such a transformation does not occur up to the decomposition of these compounds at about 1200 K. Calculation of the packing density ($p$ = total volume of oxygen ions per unit cell/volume of the unit cell) gives around 63%. This shows that the structure is based on a close packing of oxygen layers with $A$ and Te in the interstices. Second harmonic generation (SHG) measurements at room temperature gave a negative result, thus testifying that at this temperature all tested MCTO compounds most probably possess a centrosymmetric crystal structure.

### 3.2 Magnetic properties

Figure 3 shows the temperature dependence of the heat capacity and the magnetic susceptibility $\chi$=M/H for selected MCTO samples between $x$ = 0 and 3. A peak reflecting antiferromagnetic order can be seen in all heat capacity curves. The peak temperature corresponds as expected to the maximum slope of the M/H x T curves, and is indicated in red arrows in both heat capacity and magnetization data. The antiferromagnetic transition temperature $T_N$ monotonously increases from 23 K ($x$ = 0) to 42 K ($x$ = 2.7). It amounts to 32 K for the half doped $x$ = 1.5 composition. Some irreversibility above $T_N$ can be observed in the ZFC/FC curve for compositions around $x$ = 1.5 (onset near 90 K for that composition), which may indicate the presence of magnetic interaction or spurious moment. However, no feature is observed in the heat capacity curves in the vicinity of the irreversibility temperatures. On the other hand, a broad shoulder can be seen in the heat capacity below $T_N$ for all MCTO compounds with $x < 3$, which may be associated with the complex magnetic state and magnetic frustration present in the system. For some compositions we have checked that the high-temperature susceptibility data follows a Curie-Weiss behavior, $\chi(T)=C/[T-\theta_{CW}]$, by fitting to the high-temperature inverse susceptibility data [43]. The determined effective magnetic moments lie between 4.69 and 5.93 $\mu_B$, as expected from mixtures of the $Co^{2+}$ and $Mn^{2+}$ moments. The Weiss temperatures are all negative, as the main magnetic interaction is antiferromagnetic for all samples. Defining the frustration parameter $f=|\theta_{CW}/T_N|$, we can see that the frustration in interaction monotonously decreases as $x$ increases, from $x = 0$ (f ~5.2) to $x = 2.7$ (f~1.6), albeit the antiferromagnetic interaction remains large. For reference, f~2.1 for $x = 3$ (pure Co) case and f~1.84 for the $x = 2.4$ composition ($T_N = 40$ K).

### 3.3 Dielectric measurements

To gain insight into possible ferroelectric transitions, the temperature dependence of the dielectric constant ε and loss tangent of MTO and CTO single crystals were measured between 1 kHz and 1 MHz in the temperature range 300-1000K (see Figure 4). For MTO, a step is observed in both ε and loss tangent at T ≈ 860 K which was reproducible on heating and cooling.

The measurements at different frequencies reveal absence of relaxor behaviour. It was found that ε increases as frequency decreases, yet with similar overall shape and transition temperature. An additional investigation did not show a dielectric hysteresis loop up to breakdown field values, which indicates a possible antiferroelectric nature of the anomaly. The positions of the dielectric anomalies do not agree with the results obtained for ceramic samples prepared by hot pressing [34, 35], as the anomalies are registered around 500 K in that case. A possible reason for this disagreement could be related with the quality and microstructure of the ceramic samples. Concerning the ceramic solid solution series, with small Co doping ($x < 1.2$), the temperature for the anomaly in the capacitance and dissipation factor rapidly shifts to lower temperatures (420 K for $x = 1.2$) with increasing substitution. For MCTO samples with $x > 1.2$, no peak or stepwise anomaly is observed, neither in ε nor in the loss factor.

In the CTO sample two dielectric anomalies are detected: a strongly frequency-dependent feature around 600-800 K and a shoulder at 1100 K. The first one is most probably related to a diffuse phase transition. The broad maximum of the dielectric permittivity and its shifting to higher temperatures with increasing frequency indicates a relaxor-like behaviour for CTO. The second, shoulder-like anomaly in the temperature dependence of the dielectric constant at 1100 K was reproducible on heating and cooling.

### 3.4 Neutron powder diffraction at room temperature

The refined lattice parameters, atomic positions, bond lengths, and bond valence sums (see Table 3) are similar to those obtained from the XRD refinements but we were able to determine more accurately the oxygen positions due to the characteristics of the neutron scattering. No vacancies were observed in the cationic or in the anionic substructures. Accordingly, the chemical composition seems to be very close to the nominal one, and therefore oxidation state can be assumed to be 2 for both Mn and Co.

The fits to the neutron diffraction data for different compositions are shown in Fig. 5 and the structural parameters and bond lengths from the refinements are given in Tables 3 and 4. Polyhedral analysis of the different cations in MCTO is presented in Table 5. The distortion of the $TeO_6$ octahedra is less than that of the $Mn/CoO_6$ octahedra. *A*-site cations in MCTO are off-centred within the coordination polyhedra. Further distortion of the corundum structure can be driven by an increase in doping on the *A*-site. This distortion finally results in a phase transition to a structure with a more complicated stacking sequence, both along and orthogonal to the *c*-axis. Such a distorted structure has a monoclinic cell in space group *C*2/*c* ($Co_3TeO_6$) and differs in the mode of the connection of metal-oxygen polyhedra.

The crystal structure of MCTO can be derived from a close packing of strongly distorted hexagonal oxygen layers parallel to (001), with Mn/Co and two distinct Te atoms in the octahedral interstices (see Fig. 6). Both $TeO_6$ octahedra exhibit $\bar{3}$ symmetry and are fairly regular (see Table 2), with an average Te-O distance of 1.927 Å, which is in good agreement with the average Te-O distances of other tellurates with Te in octahedral coordination. Each $TeO_6$ octahedron shares edges with six $(Mn/Co)O_6$ octahedra but none with other $TeO_6$ octahedra. Each $(Mn/Co)O_6$ octahedron shares four edges with adjacent $(Mn/Co)O_6$ octahedra, one edge with a $Te(1)O_6$ and another edge with the $Te(2)O_6$ octahedron. The shared edges of the $TeO_6$ octahedra have somewhat shorter O-O distances than the non-shared edges. This could be due to the high valence of tellurium which has the tendency to stay as far as possible away from the ($Mn^{2+}$/$Co^{2+}$) cations. Each of the two crystallographically independent O atoms is coordinated by one Te and three Mn/Co atoms in a distorted tetrahedral manner. The $(Mn/Co)O_6$ octahedron is considerably distorted which is reflected by the variation of the (Mn/Co)-O distances between 2.111 (6) to 2.374 (6) Å (Table 4). The (Mn/Co)-O bond lengths, (Mn/Co)–O–(Mn/Co) angles and (Mn/Co) sublattice topology are given for different MCTO compositions in Table 4. The (Mn/Co)–O–(Mn/Co) angles vary between between 91-121°.

The position of the (Mn/Co) cations could not be extracted quantitatively from NPD for the compound with $x = 1.8$, because the corresponding contribution from this site is too small to be detected owing to a full compensation of negative (Mn) and positive (Co) scattering abilities for this content. For this compound the coordinates of (Mn/Co) cations were therefore not refined and fixed, using the results from XRPD.

### 3.5. Low-Temperature Neutron Powder Diffraction

Figure 5 shows the temperature-dependence of the NPD patterns for MCTO with varying *x*. At temperatures between 295 and 5 K, the expected decrease in the unit cell parameters is observed on cooling. At $T_N$ no anomalous features are observed in *a* and *c* parameters. No changes in the symmetry are observed in either the X-ray or neutron diffraction data from 5 K to 295 K. The patterns for all substituted samples above $T_N$ show only the peaks expected from the room-temperature crystal structure. In addition, a broad distribution of diffusive intensity is present over a notable 2θ range centered at 20°. This diffuse scattering disappears on cooling below $T_N$. Therefore, the diffuse scattering indicates the presence of short-range AFM correlations that develop above $T_N$.

Below $T_N$ the NPD patterns contain several new magnetic Bragg peaks (see Fig. 5h). Their peak positions were determined using the peak-fitting tool in the frame ''WINPLOTR software'' and the program ''K-search'' was used to find possible incommensurate propagation vectors (both programs are part of the FULLPROF software. For all doped compounds it turned out that the magnetic satellite peaks are consistent with the incommensurate antiferromagnetic (ICM AFM) structure model with $k = (0, 0, k_z)$ observed earlier for MTO [27] (see Fig. 7).

Magnetic symmetry analysis for the determination of the allowed irreducible representations of $k = [0, 0, k_z]$ in space group $R\bar{3}$ was performed using the program BASIREPS. The cation site on Wykoff position 18 *f* splits into two magnetically independent orbits, which both possess the same three allowed irreducible representations. All different combinations of basis vectors of these three irreducible representations where checked. Only one two-dimensional representation (IRep2 [27] or τ2 [44] with real and imaginary components) is able to fit the measured data. It is identical to the one already used for MTO [27] and can be used to describe the magnetic structures of all the doped MCTO compounds. For each of the two different orbits there are in theory three independent basis vectors. Coupling the two coefficients (C1 and C2) of the basis vectors lying in the hexagonal basal plane, by imposing C1 = -C2, reduced the overall number of free parameters to four (two for each orbit) and still produced very good and stable refinements. The results of the magnetic and nuclear structure refinements for different MCTO compositions are summarized in Table 6. Observed, calculated, and difference NPD patterns at 5 K for different doping concentrations are given in Figure 8.

The AFM structure adopted can be viewed as a superposition of an elliptical spiral lying within the hexagonal *ab*-plane and a longitudinal spin wave along the *c*-direction. Figure 9 displays this schematically using only one out of the 18 cation sites within the unit cell. While in Fig. 9a the longitudinal spin wave in the *c*-direction is shown, Figs. 9b and 9c represent the elliptical spiral formed within the hexagonal basal plan viewed along two different directions. A propagation vector of $k = [0, 0, 0.38]$ having a $k_z$ value different from the commensurate value of $k_z=0.5$ was used in order to facilitate the visualization of the periodicities in the *c*-direction.

Figure 10 displays the concentration-dependant variation of the coefficients for the two different orbits. It can be seen that the cation site (Mn1/Co1) adopts a purely elliptical spiral type of magnetic order in MCTO but has an increasing contribution of the longitudinal spin wave with increasing doping level *x*. At the same time the cation site (Mn2/Co2) changes from a nearly exclusively spin wave type for MTO to a dominantly elliptical spiral type in MCTO as the value of *x* increases.

Due to the high number of magnetic sites within one unit cell it is excessive to visualize the complete magnetic structure in a single figure. The magnetic structures shown in Figures 11and 12 display therefore only 3 (related through the *R*-centring condition) out of 9 symmetry related sites of the orbits (Mn1/Co1) respectively (Mn2/Co2). The decreasing predominance of the basal plane components for orbit 1 and of the *c*-component for orbit 2 with increasing *x*-values is clearly seen. The value of $k_z$ is dependent on the Co-concentration and is found to vary gradually from 0.43 for *x* = 0 towards commensurate 0.5 at *x* slightly smaller than 1.5. Surprisingly, with the further increase of the Co concentration, $k_z$ becomes incommensurate again moving to 0.516 for *x* = 2.4 (see Fig. 7). The magnetic structure having a commensurate value of $k_z = 0.5$ does not seem to be energetically favoured, since there is no sign of a lock-in transition leading to a broader stabilisation range in *x*.

The complex magnetic structure of $Mn_3TeO_6$ stems from the trigonal symmetry and the sinusoidal modulation of each of the individual components. The turn angle between two spins within one orbit is determined from the difference in the *z*-coordinate of the Mn cation and the value of the propagation vector. The spin reorientation with Co substitution is due to the strong anisotropy of $Co^{2+}$ ions [45, 46] that locks the $Mn^{2+}$ spins in the same direction and makes the collinear spin structure more favorable with increasing *x*.

Although the magnetic phase is in its nature incommensurate, there are $k_z$ values for which the period of the modulation is a rational multiple of *c*. For MTO, it appears that $k_z$ for the modulation has to a good approximation the value 3/7. This means that three helix periods add up to seven *c* lattice spacings, i.e. after seven of the original unit cells the magnetic moment takes its initial orientation. The spiral angle within the limits of the experimental error is $\Psi=2\pi \cdot k_z$ (154.8° for *x* = 0) from cell to cell. The spiral angle increases with Co doping, indicating that the structure approaches a collinear AFM model. Along the *a* and *b* axes, no incommensurate modulation is present and the repeat unit is exactly equal to the *a* parameter The resulting magnetic structure for MCTO can be described as a distorted helical structure with fluctuating (Mn/Co) moment values and directions. Although there is only one (Mn/Co) site in the $Mn_3TeO_6$ structure one does not expect perfect AFM structures due to the arrangement of the Mn-O-Mn bonds. It is important to note that in the MTO structure, three Mn atoms within a ring (at the same height in *c*-direction) have equivalent individual bonds giving the classical case of frustration in a triangle.

## 4. Discussion

Viewing the MCTO system as substituting Co for Mn in $Mn_3TeO_6$, it retains the trigonal structure of the parent compound up to more than 80% of Co substitution (possibly 90%). Using the opposing view of MCTO, i.e. substituting Mn for Co in $Co_3TeO_6$, transforms the monoclinic structure of the parent compound to the trigonal MTO structure already at less than 20% Mn substitution. This game of words highlights a key finding of this study of MCTO – the stability of the higher symmetrical structure of trigonal MTO. Elaborating further on this theme, Fig. 2 shows that the volumes of the MCTO ($x = 2.4$) and the CTO structures are quite close. Investigating the geometric consequences of a phase transformation between these structures may hint at the necessary strength of the driving mechanism behind the symmetry-breaking distortion of the CTO structure with respect to a configuration with the higher symmetry of the MTO structure. Using a set of crystallographic programs [47-49], the structures of the MCTO and CTO phases were described in conventional settings (see Fig. 6). It was possible to find an optimal mapping of both structures with a set of displacements when an atom in the MCTO structure is shifted to the closest atom of the same type in the CTO structure. Equivalent pairs of atoms between the two structures were defined and the corresponding displacements were calculated. The derived displacement components for each of the atoms are given in Table 7, in relative and absolute units. The degree of lattice distortion (**S**) - the spontaneous strain (sum of the squared eigenvalues of the strain tensor divided by 3) for the given two structures was calculated and found to be 0.0215. The maximum distance ($d_{max.}$) which shows the maximal displacement between the atomic positions of the paired atoms in this case is 1.3336 Å; associated average distance (**$d_{av}$**) was calculated as 0.2066 Å. Finally, the measure of similarity (**Δ**) (see a full definition in [50]) derived from the differences in atomic positions (weighted by the multiplicities of the sites) and the ratios of the corresponding lattice parameters of the structures was found to be 0.143. From this comparison of the crystal structures of $(Mn_{0.1}Co_{0.9})_3TeO_6$ and $Co_3TeO_6$ it is evident that the two phases are closely related and that the structure of CTO can be represented as a distorted monoclinic variant of the trigonal MTO structure of $(Mn_{0.1}Co_{0.9})_3TeO_6$.

There is no group-subgroup relation between the symmetry groups of the two structures, and instead a common subgroup or common supergroup should be investigated. Crystal-chemistry studies usually consider the common supergroup approach, especially when two structures correspond to different compounds. In our calculations we are using the common subgroup approach, in an attempt to describe the compounds constructing the so-called family or Baernighausen tree. In our case, the two structures of MTO and CTO quite nicely fit a family tree with a $R\bar{3}c$ parent phase at the top, or even a phase of higher cubic symmetry ($Pm\bar{3}n$ or $Ia\bar{3}$) aristotype symmetry. It is interesting to note that $Cu_3TeO_6$ has structure with s.g. $Ia\bar{3}$ [26]. A detailed crystallochemical analysis of $A_3TeO_6$ (*A*=Hg,Ba,Sr,Pb,Ca,Cd,Mn,Co,Ni,Cu) tellurates is in progress.

In spite of an average weakening of the exchange interaction on doping concentration (decrease of the magnitude of Weiss temperature), the transition temperature increases linearly from 23 K for $x = 0$ to 40 K for $x = 2.4$ (42 K for $x = 2.7$). This increase is accompanied and explained by considerable weakening of the magnetic frustration in the system as evidenced by the monotonous decrease of the frustration parameter from f = 5.2 for $x = 0$ to f = 1.84 for $x = 2.4$ (f = 1.6 for $x = 2.7$).

The replacement of $Mn^{2+}$ ($3d^5$) by $Co^{2+}$ ($3d^7$) in the octahedral crystal field in addition to chemical and electronic disorder also introduces the effect of a strong single-ion anisotropy of $Co^{2+}$ [51,52]. Thus, the interaction pattern becomes more complex in the triangular cycles compared to that of MTO [27]. Since all MCTO compounds adopt the MTO structure, there are two types of triangular cycles comprised of Mn/Co/Mn and Co/Mn/Co sites, stacked along the $c$ axis of the unit cell (see Fig. 13). All magnetic interactions are assumed to be AFM, although not with the same strength. Consideration of the connectivity of the (Mn/Co) cations with increasing Co substitution shows a deviation from the ideal triangular lattice of MTO. The distortions are reflected in displacements of the Co position from the initial site. The observed modifications of the elliptical spiral magnetic structure of MCTO with increasing Co substitution arise from the combination of disorder, geometrical frustration and single-ion anisotropy [51, 53].

## 5. Concluding remarks

$Mn_{3-x}Co_xTeO_6$ solid solutions preserves the trigonal corundum-related structure of MTO up to $x = 2.4$ at ambient and low temperature conditions. The lattice parameters $a$ and $c$ decrease linearly with the increase in $x$, whereas the $c/a$ ratio increases continuously with $x$.

A key feature of the structure is a close packing of strongly distorted hexagonal oxygen layers parallel to (001), with (Mn/Co) and two distinct Te atoms in the octahedral interstices. The two $TeO_6$ octahedra are fairly regular whereas the $(Mn/Co)O_6$ octahedron is considerably distorted. The substitution of Co for Mn ions changes the magnetic properties of MTO in an unpredicted way. The Neel temperature of MCTO increases linearly with the increase in $x$ but the extent of frustration, as indicated by the magnitude of the frustration parameter $\theta/T_N$, decreases with the increase in $x$. The incommensurate spin structure of MTO is preserved but modified with regard to the propagation vector and ellipticity as the Co concentration is increased. The spin anisotropy of the Co cations plays a vital role in defining the low-$T$ magnetic structures.

Substitution with Co cations thus provides a way to fine-tune the magnetic structure of MTO. The helical magnetic spin structure of MCTO and dielectric anomalies makes these compounds promising candidates for materials with spin and dipole ordering.

**Acknowledgements**

Financial support of this research from the Swedish Research Council (VR), the Göran Gustafsson Foundation and the Russian Foundation for Basic Research is gratefully acknowledged. We thank S.Yu. Stefanovich for his technical assistance with the SHG experiments. The authors are also grateful to M. I. Aroyo for stimulating discussions and assistance related to the tools hosted in the Bilbao Crystallographic Server.

**References**

[1] R. Ramamoorthy, L. Martin, Multiferroics:Synthesis, Characterization and Applications" Wiley-VCH Verlag Gmbh, 2012.
[2] S. A. Ivanov "Magnetoelectric complex metal oxides: main features of preparation, structure and properties" in "Advanced Functional Materials" Eds. B.Sanyal, O.Eriksson, Elsevier, Oxford, UK, 2012, pp 163-234.
[3] H. Schmid, J. Phys: Condens. Matter, 2008, 20, 434201.
[4] N. A. Spaldin, Topics Appl. Physics, 2007, 105, 175.
[5] "Magnetoelectric Interaction Phenomema in Crystals" Eds. A. J. Freeman, H. Schmid, Gordon and Breach, N.Y., 1975, p. 228.
[6] E. V. Wood, A.E. Austin, Inter. J. Magn., 1974, 5, 303.
[7] G. A. Smolensky, I. E. Chupis Sov. Phys.-Uspekhi, 1982, 25, 475.
[8] D. I. Khomskii, Physics, 2009, 2, 20.
[9] T. Kimura, T. Goto, H. Shintani, K. Ishizaka, T. Arima, Y. Tokura, Nature 2003, 426, 55.
[10] Y. Tokura, Science, 2006, 312, 1481.
[11] N. A.Spaldin, S. W.Cheong, R. Ramesh, Physics Today, 2010, 63(10), 38.
[12] R. Ramesh, N. A.Spaldin, Nature Materials, 2007, 6, 21.
[13] A. Filippetti, N. A. Hill, Phys. Rev. B, 2002, 65, 195120.
[14] J. Kreisel, M. Kenzelman, Europhysics News, 2009, 40, 17.
[15] C. N. R.Rao, C. R.Serrao, J. Mater. Chemistry 2007,17,4931.
[16] W. Eerenstein, N. D. Mathur, J. F. Scott, Nature, 2006, 442, 759.
[17] W. Prellier, M. P. Singh, P. Murugavel, J. Phys.: Cond. Matter, 2005, 117, R803.
[18] M. Johnsson, K. W. Tornroos, P. Lemmens, and P. Millet, Chem. Mater. 2003, 15, 68.
[19] M. Pregelj, O. Zaharko, A. Zorko, Z. Kutnjak, P. Jeglic, P. J. Brown, M. Jagodic, Z. Jaglicic, H. Berger, D. Arcon, Phys. Rev. Lett., 2009, 103, 147202.
[20] I. Zivkovic, K. Prsa, O. Zaharko, H. Berger, J. Phys. Condens. Matter, 2010, 22, 056002.
[21] R. Becker, M. Johnsson, H. Berger, Acta Crystallogr., 2006, C62, i67.
[22] J. L. Her, C. C. Chou, Y. H. Matsuda, K. Kindo, H. Berger, K. F. Tseng, C. W. Wang, W. H. Li, H. D. Yang, Phys. Rev. B, 2011, 84, 235123.
[23] M. Weil, Acta Crystallogr. 2006, E62, i244-i245.
[24] G. Caimi, L. Degiorgi, H. Berger, L. Forro, Europhys. Lett., 2006, 75, 496.
[25] K .Y. Choi, P. Lemmens, E. S. Choi, H. Berger J. Phys.: Condens. Matter 2008, 20, 505214.
[26] M. Herak, H. Berger, M. Prester, M. Miljak, I. Zivkovic, O. Milet,
D. Drobac, S. Popovic, O. Zaharko, J. Phys.:Condens. Matter, 2005, 17, 766.
[27] S. A. Ivanov, P. Nordblad, R. Mathieu, R. Tellgren, C. Ritter, N. V. Golubko, E. D. Politova, M. Weil, Mat. Res. Bull., 2011, 46, 1870.
[28] S. A. Ivanov, R. Tellgren, C. Ritter, P. Nordblad, R. Mathieu, G. André, N. V. Golubko, E. D. Politova, M. Weil, Mat. Res. Bull., 2012, 47, 63.
[29] A. B. Harris, Phys. Rev. B, 2012, 85,10.
[30] S. A. Ivanov, R. Mathieu, P. Nordblad, E. Politova, R. Tellgren, C. Ritter, V. Proidakova J. Magn. Magn. Mater., 2012, 324, 1637.
[31] G. M. Kaleva, E. D. Politova, S. A. Ivanov, A. V. Mosunov, S. Yu. Stefanovich, N. V. Sadovskaya, R. Mathieu, P. Nordblad, Inorg. Mater., 2011, 47, 1132.
[32] N. V. Golubko, V. Yu. Proidakova, G. M. Kaleva, S. A. Ivanov, A. V. Mosunov, S. Yu. Stefanovich, N. V. Sadovskaya, E. D. Politova, P. Nordblad, Bull. Russ. Acad. Science: Physics, 2010,74,724.
[33] B. Stöger, M. Weil, E. Zobetz, Z. Kristallogr. 2010, 225, 125.
[34] G. Bayer, Z. Kristallogr. 1967, 124, 131.
[35] L. I. Kosse, E. D. Politova, V. V.Chechkin, Neorg. Mater.,1982, 18, 1879.

[36] L. I. Kosse, E. D.Politova, Yu. N. Venevtsev, Neorg. Khim.,1983, 28, 1689.
[37] M. Hudl, R. Mathieu, S. A. Ivanov, M. Weil, V. Carolus, T. Lottermoser, M. Fiebig, Y. Tokunaga, Y. Taguchi, Y. Tokura, P. Nordblad, Phys. Rev. B, 2011, 84, 180404.
[38] J. Rodriguez-Carvajal, Physica B, 1993, 192, 55.
[39] T. B. Zunic, I. Vickovic, J. Appl. Phys., 1996, 29, 305.
[40] J. Rodriguez-Carvajal 2007 BASIREPS: a program for calculating irreducible representations of space groups and basis functions for axial and polar vector properties. Part of the FullProf Suite of programs available at http://www.ill.eu/sites/fullprof/. C. Ritter, Solid State Phenom., 2011, 170, 263.
[41] R. D. Shannon, Acta Crystallogr. 1976, A32, 751.
[42] I. D. Brown, Chem. Rev. 2009, 109, 6858.
[43] R. Mathieu, S. A. Ivanov, P. Nordblad, and M. Weil, Eur. Phys. J. B, 2013, 86, 361.
[44] O. V. Kovalev, in "Representations of the Crystallographic Space Groups: Irreducible representations, Induced representations and Corepresentations (2nd Ed)". Eds H. T. Stokes and D. M. Hatch, Gordon and Breach Science Publishers (London), 1993, p250.
[45] B. Forsyth, C. Wilkinson, J. Phys.: Condens. Matter, 1994, 6, 3073.
[46] N. Hollmann, Z. Hu, T. Willers, L. Bohaty, P. Becker, A. Tanaka, H. H. Hsieh, H.-J. Lin, C. T. Chen, L. H. Tjeng, Phys. Rev. B, 2010, 82, 184429.
[47] M. I. Aroyo, J. M. Perez-Mato, D. Orobengoa, E. Tasci, G. de la Flor, A. Kirov, Bulg. Chem. Commun., 2011, 43, 183.
[48] J. M. Perez-Mato, D. Orobengoa, M. I. Aroy, Acta. Crystallogr., 2010, A66, 558.
[49] M. I. Aroyo, J. M. Perez-Mato, C. Capillas, E. Kroumova, S. Ivantchev, G. Madariaga, A. Kirov, H. Wondratschek , Z. Kristallogr., 2006, 221, 15
[50] G. Bergerhoff, M. Berndt, K. Brandenburg, T. Degen, Acta Crystallogr. 1999, B55, 147.
[51] R.L. Carlin, "Magnetochemistry" (Springer Verlag, Berlin) (1986).
[52] A. Abragam, M. H. L. Pryce, Proc. Roy. Soc. 1951, A206, 173.
[53] M. F. Collins, O. A. Petrenko Can. J. Phys., 1997, 75, 605.

# Figure captions

**Figure 1** (a) XRPD patterns of $Mn_{3-x}Co_xTeO_6$ (MCTO) compounds with different $x$, and Rietveld fits for $x = 0$ (b), $x = 0.3$ (c), $x = 0.9$ (d), $x = 1.5$ (e) and $x = 2.4$ (f).

**Figure 2** Refined lattice parameters ($a,c$) and cell volumes ($V$) of MCTO powder samples as a function of $x$. The solid lines are drawn through the lattice parameter points as a guide for the eyes.

**Figure 3** Temperature and concentration dependence of (left) the heat capacity C (plotted as C/T) and (right) the ZFC/FC susceptibility M/H of MCTO. Red arrows denote the respective antiferromagnetic transition temperatures. In case of $x = 0$ and 3 (pure $M$ = Mn and $M$ = Co compounds), results for single crystals are shown, including different magnetic field orientations.

**Figure 4** Dielectric characteristic curves of CTO (a,c) and MTO (b,d) at different frequencies: (1: 0.1, 2: 1, 3: 10, 4: 100, 5: 1000 kHz).

**Figure 5** (a-g) The temperature-dependence of NPD pattern for different Co molar fraction: $x = 0$ (a), 0.6 (b), 0.9 (c),1.2 (d),1.5 (e),1.8 (f) and 2.4 (g). (h) Magnetic reflections (indicated by asterisks) of MCTO for different Co concentrations.

**Figure 6** Polyhedral representation of the crystal structures of MTO (a) and CTO (b).

**Figure 7** Variation of the propagation vector $k_z$ of MCTO with Co molar fraction $x$.

**Figure 8** Neutron powder diffraction profiles of MCTO for different concentrations: x = 0 (a), x = 0.9 (b), x = 1.5 (c) and x = 2.4 (d). The crosses and the lines indicate experimental and calculated intensities, respectively. The solid line at the bottom of each subfigure is the difference between the two intensities. The vertical marks indicate Bragg-peak positions of the nuclear (upper row) and the magnetic (lower row) phases.

**Figure 9** Schematic view of the spin structure using only one out of the 18 cation sites within the unit cell: a) the longitudinal spin wave in $c$-direction, b) and c) represent the elliptical spiral formed within the hexagonal basal plan viewed along two different directions.

**Figure 10** Variation of the refined values of the coefficients $C_1$ and $C_3$ of the magnetic phase of MCTO at 5 K with Co molar fraction $x$.

**Figure 11** Schematic representation of the incommensurate magnetic structure for different concentrations of MCTO for three cells along the direction of the propagation vector ($Mn_1$ orbit).

**Figure 12** Schematic representation of the incommensurate magnetic structure for different concentrations of MCTO for three cells along the direction of the propagation vector ($Mn_2$ orbit).

**Figure 13** Triangle configurations of magnetic cations for MTO (a) and CTO (b) lattices.

**Table 1** Results of the Rietveld refinements of the crystal structure of the MCTO samples at room temperature using X-ray powder diffraction data (s.g. R-3, A=Mn/Co). The lattice parameter standard deviations are smaller than $3 \times 10^{-4}$ Å. Standard deviations of atomic positions and occupation factors were less than 0.003 Å and 0.02, respectively. Occupancies obtained from EDS analysis are included for comparison.

| Phase | | $x = 0$ | $x = 0.3$ | $x = 0.6$ | $x = 0.9$ | $x = 1.2$ | $x = 1.5$ | $x = 1.8$ | $x = 2.1$ | $x = 2.4$ |
|---|---|---|---|---|---|---|---|---|---|---|
| | $a$, Å | 8.8675 | 8.8447 | 8.8155 | 8.7880 | 8.7597 | 8.7270 | 8.7061 | 8.6873 | 8.6601 |
| | $c$, Å | 10.6731 | 10.6559 | 10.6350 | 10.6137 | 10.5908 | 10.5644 | 10.5514 | 10.5354 | 10.5150 |
| | $c/a$ | 1.2036 | 1.2048 | 1.2064 | 1.2078 | 1.2090 | 1.2105 | 1.2120 | 1.2127 | 1.2142 |
| Mn/Co (EDS) | | 2.98 | 2.72/0.27 | 2.38/0.63 | 2.06./0.92 | 1.82/1.18 | 1.52/1.46 | 1.17/1.82 | 0.88/2.13 | 0.59/2.38 |
| Te (EDS) | | 1.02 | 1.01 | 0.99 | 1.02 | 1.00 | 1.02 | 1.01 | 0.99 | 1.03 |
| A | $n$ | 2.98/0 | 2.68/0.32 | 2.37/0.63 | 2.13/0.87 | 1.83/1.17 | 1.47/1.53 | 1.23/1.77 | 0.93/2.07 | 0.62/2.38 |
| | $x/a$ | 0.0385 | 0.0391 | 0.0396 | 0.0402 | 0.0406 | 0.0410 | 0.0408 | 0.0417 | 0.0421 |
| | $y/b$ | 0.2640 | 0.2640 | 0.2636 | 0.2640 | 0.2639 | 0.2631 | 0.2622 | 0.2629 | 0.2627 |
| | $z/c$ | 0.2128 | 0.2128 | 0.2129 | 0.2127 | 0.2123 | 0.2122 | 0.2118 | 0.2117 | 0.2111 |
| | $B_{eq}$ /Å$^2$ | 0.69(3) | 0.67(3) | 0.65(3) | 0.62(3) | 0.59(3) | 0.56(3) | 0.53(3) | 0.51(3) | 0.48(3) |
| Te1 | $n$ | 1.02(2) | 0.98(2) | 0.99(2) | 1.01(2) | 1.00(2) | 0.98(2) | 0.99(2) | 1.01(2) | 1.02(2) |
| | $x/a$ | 0 | 0 | 0 | 0 | 0 | 0 | 0 | 0 | 0 |
| | $y/b$ | 0 | 0 | 0 | 0 | 0 | 0 | 0 | 0 | 0 |
| | $z/c$ | 1/2 | 1/2 | 1/2 | 1/2 | 1/2 | 1/2 | 1/2 | 1/2 | 1/2 |
| | $B_{eq}$ /Å$^2$ | 0.40(2) | 0.38(2) | 0.38(2) | 0.37(2) | 0.35(2) | 0.36(2) | 0.34(2) | 0.35(2) | 0.33(2) |
| Te2 | $n$ | 1.01(2) | 0.98(2) | 0.99(2) | 1.01(2) | 1.00(2) | 1.02(2) | 0.99(2) | 1.01(2) | 1.02(2) |
| | $x/a$ | 0 | 0 | 0 | 0 | 0 | 0 | 0 | 0 | 0 |
| | $y/b$ | 0 | 0 | 0 | 0 | 0 | 0 | 0 | 0 | 0 |
| | $z/c$ | 0 | 0 | 0 | 0 | 0 | 0 | 0 | 0 | 0 |
| | $B_{eq}$ /Å$^2$ | 0.38(2) | 0.39(2) | 0.37(2) | 0.35(2) | 0.36(2) | 0.34(2) | 0.35(2) | 0.33(2) | 0.34(2) |
| O1 | $n$ | 1.01(3) | 0.99(3) | 1.03(3) | 0.98(3) | 1.01(3) | 0.97(3) | 0.98(3) | 0.97(3) | 1.02(3) |
| | $x/a$ | 0.0307 | 0.0292 | 0.0319 | 0.0303 | 0.0309 | 0.0328 | 0.0299 | 0.0332 | 0.0357 |
| | $y/b$ | 0.1963 | 0.2014 | 0.2027 | 0.2020 | 0.2015 | 0.2029 | 0.2056 | 0.2051 | 0.2053 |
| | $z/c$ | 0.4028 | 0.3948 | 0.3951 | 0.3952 | 0.3960 | 0.3956 | 0.3946 | 0.3906 | 0.3891 |
| | $B_{eq}$ /Å$^2$ | 0.84(2) | 0.86(2) | 0.82(2) | 0.79(2) | 0.83(3) | 0.81(3) | 0.76(3) | 0.78(3) | 0.74(3) |
| O2 | $n$ | 0.98(2) | 1.02(3) | 0.97(2) | 1.03(2) | 0.99(3) | 0.98(3) | 0.97(3) | 1.01(2) | 0.98(3) |
| | $x/a$ | 0.1828 | 0.1846 | 0.1862 | 0.1852 | 0.1877 | 0.1889 | 0.1891 | 0.1869 | 0.1859 |
| | $y/b$ | 0.1562 | 0.1603 | 0.1602 | 0.1618 | 0.1635 | 0.1631 | 0.1629 | 0.1629 | 0.1628 |
| | $z/c$ | 0.1105 | 0.1166 | 0.1160 | 0.1167 | 0.1172 | 0.1175 | 0.1172 | 0.1165 | 0.1149 |
| | $B_{eq}$ /Å$^2$ | 0.69(2) | 0.70(2) | 0.67(2) | 0.68(2) | 0.65(2) | 0.63(2) | 0.65(2) | 0.63(2) | 0.62(2) |
| | $R_p$ | 4.04 | 4.21 | 4.11 | 4.23 | 4.52 | 4.64 | 4.76 | 4.82 | 5.07 |
| | $R_{wp}$ | 5.32 | 5.57 | 5.26 | 5.47 | 5.57 | 5.83 | 6.02 | 5.93 | 6.64 |
| | $R_b$ | 3.87 | 4.22 | 4.17 | 4.21 | 4.43 | 4.57 | 4.71 | 4.82 | 5.02 |
| | $\chi^2$ | 1.19 | 1.23 | 1.09 | 1.30 | 1.16 | 1.22 | 1.14 | 1.30 | 1.19 |

**Table 2** Selected bond lengths and angles from XRPD powder refinements of MCTO samples at room temperature ($A$ = Mn/Co).

| Phase | | $x = 0$ | $x = 0.3$ | $x = 0.6$ | $x = 0.9$ | $x = 1.2$ | $x = 1.5$ | $x = 1.8$ | $x = 2.1$ | $x = 2.4$ |
|---|---|---|---|---|---|---|---|---|---|---|
| $A$ | O1 | 2.106(2) | 2.019(3) | 2.016(2) | 2.014(2) | 2.013(3) | 2.011(2) | 1.991(3) | 1.953(2) | 1.939(3) |
| | O1 | 2.395(2) | 2.391(3) | 2.389(2) | 2.383(2) | 2.379(3) | 2.374(2) | 2.358(2) | 2.393(2) | 2.406(3) |
| | O1 | 2.210(2) | 2.203(3) | 2.190(2) | 2.189(2) | 2.179(3) | 2.155(2) | 2.164(3) | 2.159(3) | 2.146(3) |
| | O2 | 2.228(2) | 2.176(2) | 2.171(3) | 2.142(3) | 2.134(2) | 2.128(2) | 2.122(2) | 2.099(2) | 2.089(2) |
| | O2 | 2.230(3) | 2.217(2) | 2.214(3) | 2.213(3) | 2.212(3) | 2.187(3) | 2.176(3) | 2.191(3) | 2.189(2) |
| | O2 | 2.130(2) | 2.100(3) | 2.089(3) | 2.087(3) | 2.062(3) | 2.058(2) | 2.057(2) | 2.072(2) | 2.079(3) |
| Te1 | O1 x 6 | 1.925(2) | 2.002(2) | 1.986(2) | 1.981(2) | 1.963(2) | 1.964(2) | 1.992(2) | 1.996(2) | 1.998(2) |
| Te2 | O2 x 6 | 1.921(2) | 1.979(2) | 1.983(2) | 1.980(2) | 1.999(2) | 1.998(2) | 1.991(2) | 1.975(2) | 1.952(2) |
| $A$-O1-$A$ | | 117.6(4) | 120.8(5) | 120.4(4) | 120.8(3) | 120.4(5) | 120.4(4) | 122.3(4) | 122.0(5) | 121.8(4) |
| | | 97.6(3) | 99.5(4) | 99.4(3) | 99.1(3) | 98.8(4) | 99.6(3) | 100.1(4) | 100.7(3) | 101.6(3) |
| | | 93.8(4) | 92.7(3) | 93.0(4) | 92.9(4) | 92.9(3) | 93.0(4) | 93.2(5) | 91.9(4) | 91.4(4) |
| $A$-O2-$A$ | | 116.6(4) | 118.3(4) | 118.5(5) | 118.6(4) | 118.8(4) | 118.6(5) | 118.4(4) | 118.5(4) | 118.4(4) |
| | | 100.4(3) | 103.0(3) | 102.8(3) | 103.3(4) | 103.5(4) | 103.8(4) | 103.7(5) | 103.5(4) | 103.4(4) |
| | | 98.5(4) | 99.5(4) | 100.4(5) | 99.9(4) | 100.8(5) | 101.5(4) | 101.2(4) | 100.8(4) | 100.6(4) |
| $A$-$A$ | | 3.305(3) | 3.302(3) | 3.304(3) | 3.302(3) | 3.296(3) | 3.288(3) | 3.286(3) | 3.283(3) | 3.277(3) |
| | | 3.350(3) | 3.342(3) | 3.331(3) | 3.316(3) | 3.303(3) | 3.294(3) | 3.287(3) | 3.276(3) | 3.264(3) |
| | | 3.241(3) | 3.227(3) | 3.209(3) | 3.199(3) | 3.192(3) | 3.187(3) | 3.181(3) | 3.169(3) | 3.165(3) |

**Table 3** Summary of structural refinement results of MCTO samples using NPD data ($A$ = Mn/Co). The lattice parameter standard deviations are smaller than $5\times10^{-4}$ Å. Standard deviations of atomic positions and occupation factors are less than 0.001 Å and 0.02, respectively.

| Phase | | x =0 | | x=0.6 | | x=0.9 | | x=1.2 | | x=1.5 | | x=1.8 | | x=2.4 | |
|---|---|---|---|---|---|---|---|---|---|---|---|---|---|---|---|
| T,K | | 1.5 | 295 | 5 | 295 | 5 | 295 | 5 | 295 | 5 | 295 | 5 | 295 | 5 | 295 |
| $a$,Å | | 8.8516 | 8.8679 | 8.7980 | 8.8138 | 8.7721 | 8.7882 | 8.7441 | 8.7602 | 8.7123 | 8.7288 | 8.6910 | 8.7072 | 8.6441 | 8.6589 |
| $c$,Å | | 10.650 | 10.673 | 10.612 | 10.632 | 10.593 | 10.614 | 10.571 | 10.591 | 10.547 | 10.566 | 10.530 | 10.548 | 10.495 | 10.512 |
| $A$ | n | 0.98 | | 2.39/0.61 | | 2.12/0.88 | | 1.78/1.22 | | 1.52/1.48 | | 1.22/1.78 | | 0.62/2.38 | |
| | x/a | 0.0397 | 0.0380 | 0.0388 | 0.0384 | 0.0394 | 0.0380 | 0.0400 | 0.0340 | 0.0395 | 0.0409 | 0.0404 | 0.0408 | 0.0377 | 0.0421 |
| | y/b | 0.2660 | 0.2649 | 0.2650 | 0.2663 | 0.2647 | 0.2657 | 0.2657 | 0.2680 | 0.2641 | 0.2631 | 0.2700 | 0.2639 | 0.2689 | 0.2627 |
| | z/c | 0.2131 | 0.2128 | 0.2118 | 0.2132 | 0.2125 | 0.2137 | 0.2150 | 0.2095 | 0.2109 | 0.2122 | 0.2271 | 0.2123 | 0.2200 | 0.2135 |
| | $B$,Å$^2$ | 0.21 | 0.51 | 0.20 | 0.49 | 0.22 | 0.47 | 0.23 | 0.46 | 0.21 | 0.45 | 0.23 | 0.43 | 0.20 | 0.42 |
| Te1 | n | 1.01 | | 1.02 | | 0.99 | | 1.01 | | 0.98 | | 1.02 | | 1.01 | |
| | $B$,Å$^2$ | 0.46 | 0.71 | 0.33 | 0.66 | 0.35 | 0.36 | 0.31 | 0.37 | 0.29 | 0.35 | 0.25 | 0.34 | 0.27 | 0.41 |
| Te2 | n | 1.02 | | 0.98 | | 0.99 | | 1.02 | | 1.01 | | 1.02 | | 0.99 | |
| | $B$,Å$^2$ | 0.08 | 0.32 | 0.13 | 0.33 | 0.11 | 0.29 | 0.12 | 0.31 | 0.09 | 0.27 | 0.11 | 0.26 | 0.13 | 0.24 |
| O1 | n | 1.02 | | 1.01 | | 0.98 | | 1.01 | | 1.00 | | 0.98 | | 0.99 | |
| | x/a | 0.0292 | 0.0300 | 0.0300 | 0.0301 | 0.0309 | 0.0306 | 0.0304 | 0.0287 | 0.0300 | 0.0297 | 0.0307 | 0.0314 | 0.0290 | 0.0295 |
| | y/b | 0.1967 | 0.1961 | 0.1981 | 0.1979 | 0.1989 | 0.1983 | 0.1989 | 0.1976 | 0.1992 | 0.1986 | 0.2003 | 0.1998 | 0.2003 | 0.2002 |
| | z/c | 0.4031 | 0.4030 | 0.4026 | 0.4028 | 0.4025 | 0.4026 | 0.4025 | 0.4030 | 0.4032 | 0.4031 | 0.4024 | 0.4025 | 0.4024 | 0.4024 |
| | $B$,Å$^2$ | 0.16 | 0.51 | 0.23 | 0.63 | 0.19 | 0.59 | 0.21 | 0.56 | 0.24 | 0.58 | 0.20 | 0.63 | 0.25 | 0.66 |
| O2 | n | 0.99 | | 1.01 | | 1.00 | | 1.02 | | 0.98 | | 1.01 | | 1.02 | |
| | x/a | 0.1836 | 0.1830 | 0.1831 | 0.1833 | 0.1830 | 0.1838 | 0.1841 | 0.1842 | 0.1862 | 0.1857 | 0.1846 | 0.1850 | 0.1845 | 0.1864 |
| | y/b | 0.1573 | 0.1561 | 0.1572 | 0.1561 | 0.1581 | 0.1575 | 0.1588 | 0.1562 | 0.1602 | 0.1599 | 0.1592 | 0.1599 | 0.1591 | 0.1607 |
| | z/c | 0.1110 | 0.1113 | 0.1116 | 0.1114 | 0.1120 | 0.1116 | 0.1120 | 0.1117 | 0.1128 | 0.1123 | 0.1128 | 0.1124 | 0.1126 | 0.1131 |
| | $B$,Å$^2$ | 0.20 | 0.55 | 0.31 | 0.59 | 0.33 | 0.65 | 0.35 | 0.69 | 0.41 | 0.68 | 0.44 | 0.72 | 0.46 | 0.76 |
| $Rp$ | | 4.65 | 3.16 | 5.31 | 4.46 | 4.97 | 3.78 | 5.08 | 5.49 | 5.59 | 5.13 | 5.25 | 5.49 | 5.39 | 5.25 |
| $Rwp$ | | 6.18 | 4.21 | 7.25 | 5.67 | 6.69 | 5.04 | 6.22 | 6.47 | 6.53 | 6.43 | 6.36 | 6.47 | 6.61 | 6.44 |
| $RB$ | | 2.57 | 2.61 | 3.32 | 2.95 | 2.78 | 2.75 | 3.06 | 2.82 | 3.53 | 3.14 | 3.11 | 2.98 | 3.26 | 3.07 |
| $Rm$ | | 7.68 | | 8.3 | | 8.5 | | 8.2 | | 8.1 | | 7.1 | | 8.6 | |
| $\chi^2$ | | 2.23 | 1.65 | 1.93 | 1.77 | 2.18 | 2.03 | 1.86 | 2.11 | 2.09 | 1.79 | 1.96 | 1.99 | 2.04 | 2.11 |

**Table 4** Selected bond lengths from NPD powder refinements of MCTO samples at different temperatures ($A$ = Mn/Co)

| Phase | | x=0 | | x=0.6 | | x=0.9 | | x=1.2 | | x=1.5 | | x=1.8 | | x=2.4 | |
|---|---|---|---|---|---|---|---|---|---|---|---|---|---|---|---|
| T,K | | 10 | 295 | 5 | 295 | 5 | 295 | 5 | 295 | 5 | 295 | 5 | 295 | 5 | 295 |
| $A$ | O1 | 2.103 | 2.111 | 2.099 | 2.093 | 2.085 | 2.082 | 2.056 | 2.134 | 2.096 | 2.081 | 2.036 | 2.110 | 1.994 | 2.047 |
| | O1 | 2.357 | 2.374 | 2.342 | 2.349 | 2.346 | 2.352 | 2.348 | 2.290 | 2.312 | 2.339 | 2.344 | 2.291 | 2.320 | 2.326 |
| | O1 | 2.191 | 2.206 | 2.183 | 2.184 | 2.165 | 2.174 | 2.148 | 2.208 | 2.158 | 2.154 | 2.130 | 2.138 | 2.123 | 2.130 |
| | O2 | 2.226 | 2.238 | 2.201 | 2.237 | 2.185 | 2.222 | 2.197 | 2.248 | 2.164 | 2.159 | 2.170 | 2.166 | 2.134 | 2.139 |
| | O2 | 2.242 | 2.225 | 2.214 | 2.221 | 2.217 | 2.222 | 2.230 | 2.183 | 2.190 | 2.198 | 2.211 | 2.209 | 2.189 | 2.195 |
| | O2 | 2.117 | 2.118 | 2.105 | 2.099 | 2.103 | 2.086 | 2.087 | 2.037 | 2.070 | 2.099 | 2.083 | 2.065 | 2.021 | 2.083 |
| Te1 | O1 | 1.927 | 1.925 | 1.928 | 1.929 | 1.927 | 1.926 | 1.922 | 1.918 | 1.916 | 1.917 | 1.921 | 1.919 | 1.917 | 1.919 |
| Te2 | O2 | 1.926 | 1.929 | 1.919 | 1.919 | 1.919 | 1.921 | 1.920 | 1.915 | 1.930 | 1.926 | 1.919 | 1.923 | 1.908 | 1.926 |
| $A$-O1-$A$ | | 118.2 | 117.5 | 118.3 | 118.2 | 118.7 | 118.3 | 119.2 | 116.9 | 118.9 | 119.3 | 120.0 | 118.5 | 120.1 | 120.7 |
| | | 97.0 | 97.5 | 98.1 | 97.6 | 98.1 | 97.8 | 97.2 | 99.2 | 98.3 | 97.8 | 97.2 | 99.02 | 96.1 | 97.5 |
| | | 94.1 | 93.8 | 94.0 | 94.2 | 94.2 | 94.2 | 94.5 | 93.5 | 94.3 | 94.3 | 94.8 | 94.1 | 95.3 | 94.6 |
| $A$-O2-$A$ | | 117.0 | 117.3 | 117.6 | 117.3 | 117.2 | 116.9 | 116.2 | 119.9 | 117.5 | 116.6 | 115.7 | 119.1 | 115.2 | 115.8 |
| | | 100.1 | 100.3 | 100.5 | 100.6 | 100.9 | 100.6 | 101.0 | 99.7 | 101.5 | 101.4 | 101.6 | 100.1 | 101.1 | 102.0 |
| | | 99.0 | 98.7 | 98.6 | 98.7 | 98.7 | 98.8 | 99.4 | 97.6 | 99.4 | 99.7 | 99.9 | 99.1 | 99.7 | 100.4 |
| $A$-$A$ | | 3.316 | 3.295 | 3.275 | 3.279 | 3.279 | 3.272 | 3.293 | 3.177 | 3.210 | 3.285 | 3.287 | 3.254 | 3.258 | 3.285 |
| | | 3.331 | 3.345 | 3.311 | 3.322 | 3.307 | 3.317 | 3.305 | 3.277 | 3.362 | 3.295 | 3.297 | 3.244 | 3.289 | 3.278 |
| | | 3.219 | 3.245 | 3.234 | 3.219 | 3.210 | 3.208 | 3.154 | 3.307 | 3.122 | 3.191 | 3.126 | 3.230 | 3.063 | 3.140 |

**Table 5** Polyhedral analysis of MCTO samples at 295K (*x*-concentration of Co, δ – cation shift from centroid, ξ- average bond length and its ranging limits, *V*- polyhedral volume, Δ - polyhedral volume distortion).

| Cation | *x* | δ(Å) | ξ (Å) | *V*(Å$^3$) | Δ | Valence |
|---|---|---|---|---|---|---|
| (Mn/Co) (c.n.=6) | 0 | 0.074 | 2.212+/-0.096 | 12.80(13) | 0.104 | 1.97 |
| | 0.3 | 0.053 | 2.185+/-0.131 | 12.43(13) | 0.099 | 2.17 |
| | 0.6 | 0.051 | 2.178+/-0.135 | 12.30(13) | 0.098 | 2.21 |
| | 0.9 | 0.047 | 2.171+/-0.130 | 12.19(12) | 0.097 | 2.23 |
| | 1.2 | 0.045 | 2.163+/-0.127 | 12.05(12) | 0.096 | 2.26 |
| | 1.5 | 0.042 | 2.153+/-0.131 | 11.88(12) | 0.095 | 2.22 |
| | 1.8 | 0.039 | 2.144+/-0.127 | 11.79(12) | 0.094 | 2.19 |
| | 2.1 | 0.037 | 2.143+/-0.151 | 11.78(12) | 0.093 | 2.24 |
| | 2.4 | 0.034 | 2.140+/-0.160 | 11.76(12) | 0.092 | 2.23 |
| Te$_1$ (c.n.=6) | 0 | 0 | 1.925+/-0.004 | 9.44(11) | 0.007 | 5.89 |
| | 0.3 | 0 | 2.009+/-0.004 | 10.79(12) | 0.005 | 5.68 |
| | 0.6 | 0 | 1.997+/-0.004 | 10.71(12) | 0.004 | 5.74 |
| | 0.9 | 0 | 1.981+/-0.004 | 10.60(12) | 0.003 | 5.84 |
| | 1.2 | 0 | 1.982+/-0.004 | 10.36(11) | 0.002 | 5.75 |
| | 1.5 | 0 | 1.994+/-0.004 | 10.34(11) | 0.001 | 5.81 |
| | 1.8 | 0 | 2.005+/-0.005 | 10.54(11) | 0.001 | 5.71 |
| | 2.1 | 0 | 1.999+/-0.005 | 10.74(12) | 0 | 5.69 |
| | 2.4 | 0 | 1.985+/-0.005 | 10.66(12) | 0 | 5.71 |
| Te$_2$ (c.n.=6) | 0 | 0 | 1.929+/-0.005 | 9.49(11) | 0.007 | 5.83 |
| | 0.3 | 0 | 1.973+/-0.006 | 10.11(11) | 0.006 | 5.69 |
| | 0.6 | 0 | 1.976+/-0.006 | 10.19(11) | 0.008 | 5.73 |
| | 0.9 | 0 | 1.979+/-0.005 | 10.25(11) | 0.009 | 5.66 |
| | 1.2 | 0 | 1.982+/-0.005 | 10.32(11) | 0.010 | 5.79 |
| | 1.5 | 0 | 1.984+/-0.005 | 10.31(11) | 0.011 | 5.62 |
| | 1.8 | 0 | 1.979+/-0.006 | 10.23(11) | 0.012 | 5.68 |
| | 2.1 | 0 | 1.968+/-0.005 | 9.95(11) | 0.011 | 5.75 |
| | 2.4 | 0 | 1.965+/-0.006 | 9.66(11) | 0.009 | 5.63 |

**Table 6** Refined parameters of the magnetic structure of MCTO at 5K (coefficients C1 and C2 are not orthogonal to each other but span 120°). Mn1 on $x, y, z$ (0.039, 0.265, 0.213) belongs to orbit 1, Mn2 on $-x, -y, -z$ (0.961, 0.735, 0.787) belongs to orbit 2. The phase between two orbits is around 0.01-0.10(1).

| Orbit | $x$ | C1 | C3 | $k_z$ | $\mu_{min}$ ($\mu_B$) | $\mu_{max}$ ($\mu_B$) |
|---|---|---|---|---|---|---|
| **Mn1** | 0 | 5.08(8) | 0 | 0.4302(1) | 3.6(2) | 6.2(2) |
| | 0.6 | 3.95(7) | -1.43(9) | 0.4664(2) | 3.1(2) | 4.9(2) |
| | 0.9 | 3.74(8) | -1.92(9) | 0.4781(3) | 3.0(2) | 4.7(2) |
| | 1.2 | 3.69(7) | -2.04(8) | 0.4893(3) | 2.9(2) | 4.5(2) |
| | 1.5 | 3.54(8) | -2.17(8) | 0.5007(2) | 2.8(2) | 4.3(2) |
| | 1.8 | 3.40(8) | -2.30(8) | 0.5069(2) | 2.7(2) | 4.1(2) |
| | 2.4 | 3.13(7) | -2.53(9) | 0.5157(2) | 2.6(2) | 3.8(2) |
| **Mn2** | 0 | 1.37(6) | 4.76(9) | 0.4302(1) | 1.7(2) | 5.0(2) |
| | 0.6 | 1.87(7) | 3.23(8) | 0.4664(2) | 2.2(2) | 3.9(2) |
| | 0.9 | 2.18(7) | 2.89(9) | 0.4781(3) | 2.3(2) | 3.7(2) |
| | 1.2 | 2.48(6) | 2.45(8) | 0.4893(3) | 2.3(2) | 3.6(2) |
| | 1.5 | 2.80(6) | 2.03(8) | 0.5007(2) | 2.4(2) | 3.5(2) |
| | 1.8 | 3.14(7) | 1.71(9) | 0.5069(2) | 2.5(2) | 3.4(2) |
| | 2.4 | 3.50(6) | 1.43(8) | 0.5157(2) | 2.6(2) | 3.7(2) |

**Table 7** Comparison of the crystal structures of $Mn_{0.6}Co_{2.4}TeO_6$ and $Co_3TeO_6$ represented in the same symmetry $P\bar{1}$. $\Delta x$, $\Delta y$, $\Delta z$ are given in relative units. $|\Delta|$ is the absolute distance between atom pairings given in Å

| $Co_3TeO_6$ a=8.6242 b=8.6242 c=10.3589A $\alpha$=94.15, $\beta$=85.85, $\gamma$=118.34 | | | | $Mn_{0.6}Co_{2.4}TeO_6$ a=8.6590 b=8.6590 c=10.5122A $\alpha$=90.00, $\beta$=90.00, $\gamma$=120.00 | | | | Atomic displacements | | | |
|---|---|---|---|---|---|---|---|---|---|---|---|
| Atom | x | y | z | Atom | x | y | z | $\Delta x$ | $\Delta y$ | $\Delta z$ | $|\Delta|$ |
| $Te_{11}$ | 0.5000 | 0.5000 | 0.5000 | $Te_{21}$ | 0.5000 | 0.5000 | 0.5000 | 0.0000 | 0.0000 | 0.0000 | 0.0000 |
| $Te_{12}$ | 0.5000 | 0.5000 | 0 | $Te_{11}$ | 0.5000 | 0.5000 | 0 | 0.0000 | 0.0000 | 0.0000 | 0.0000 |
| $Te_{21}$ | 0.8390 | 0.1632 | 0.2998 | $Te_{12}$ | 0.8333 | 0.1667 | 0.3333 | -0.0057 | 0.0035 | 0.0335 | 0.3484 |
| $Te_{22}$ | 0.1632 | 0.8390 | 0.2002 | $Te_{22}$ | 0.1667 | 0.8333 | 0.1667 | 0.0035 | -0.0057 | -0.0335 | 0.3484 |
| $Co_1$ | 0.6848 | 0.6848 | 0.2500 | $Mn_{12}$ | 0.7186 | 0.7625 | 0.2865 | 0.0338 | 0.0777 | 0.0365 | 0.6863 |
| $Co_{21}$ | 0.4995 | 0.2117 | 0.2321 | $Mn_{13}$ | 0.5438 | 0.2813 | 0.2865 | 0.0444 | 0.0697 | 0.0544 | 0.7674 |
| $Co_{22}$ | 0.2117 | 0.4995 | 0.2679 | $Mn_{11}$ | 0.2374 | 0.4561 | 0.2865 | 0.0258 | -0.0434 | 0.0186 | 0.5675 |
| $Co_{31}$ | 0.1293 | 0.1759 | 0.0419 | $Mn_{14}$ | 0.0958 | 0.2105 | 0.0468 | -0.0334 | 0.0346 | 0.0049 | 0.5025 |
| $Co_{32}$ | 0.1759 | 0.1293 | 0.4581 | $Mn_{19}$ | 0.1227 | 0.0519 | 0.3801 | -0.0531 | -0.0773 | -0.0779 | 0.9959 |
| $Co_{41}$ | 0.6280 | 0.9582 | 0.0573 | $Mn_{15}$ | 0.6146 | 0.9041 | 0.0468 | -0.0133 | -0.0541 | -0.0105 | 0.4312 |
| $Co_{42}$ | 0.9582 | 0.6280 | 0.4427 | $Mn_{18}$ | 0.9480 | 0.5708 | 0.3801 | -0.0102 | -0.0572 | -0.0625 | 0.7691 |
| $Co_{51}$ | 0.5616 | 0.1626 | 0.5699 | $Mn_{17}$ | 0.5708 | 0.1227 | 0.6198 | 0.0092 | -0.0398 | 0.0499 | 0.6703 |
| $Co_{52}$ | 0.1626 | 0.5616 | 0.9301 | $Mn_{16}$ | 0.2105 | 0.6146 | 0.9531 | 0.0479 | 0.0531 | 0.0231 | 0.5064 |
| $O_{11}$ | 0.4140 | 0.2836 | 0.5703 | $O_{22}$ | 0.5256 | 0.3392 | 0.6130 | 0.1117 | 0.0557 | 0.0428 | 0.9728 |
| $O_{12}$ | 0.2836 | 0.4140 | 0.9297 | $O_{12}$ | 0.3292 | 0.2998 | 0.9023 | 0.0457 | -0.1142 | -0.0273 | 1.2316 |
| $O_{21}$ | 0.7404 | 0.9432 | 0.1995 | $O_{15}$ | 0.6625 | 0.9664 | 0.2357 | -0.0778 | 0.0233 | 0.0362 | 0.8437 |
| $O_{22}$ | 0.9432 | 0.7404 | 0.3005 | $O_{29}$ | 0.9802 | 0.8076 | 0.2797 | 0.0371 | 0.0673 | -0.0208 | 0.5630 |
| $O_{31}$ | 0.0626 | 0.2598 | 0.1915 | $O_{14}$ | 0.0335 | 0.1961 | 0.2357 | -0.0291 | -0.0637 | 0.0442 | 0.6807 |
| $O_{32}$ | 0.2598 | 0.0626 | 0.3085 | $O_{27}$ | 0.3273 | 0.0197 | 0.2797 | 0.0676 | -0.0429 | -0.0288 | 0.8537 |
| $O_{41}$ | 0.7831 | 0.2643 | 0.6618 | $O_{28}$ | 0.8076 | 0.3273 | 0.7202 | 0.0246 | 0.0631 | 0.0585 | 0.7544 |
| $O_{42}$ | 0.2643 | 0.7831 | 0.8382 | $O_{16}$ | 0.1961 | 0.6625 | 0.7642 | -0.0682 | -0.1205 | -0.0739 | 1.1755 |
| $O_{51}$ | 0.5775 | 0.4403 | 0.3347 | $O_{23}$ | 0.6864 | 0.5256 | 0.3869 | 0.1089 | 0.0854 | 0.0522 | 1.0374 |
| $O_{52}$ | 0.4403 | 0.5775 | 0.1653 | $O_{11}$ | 0.2998 | 0.4705 | 0.0976 | -0.1405 | -0.1070 | -0.0677 | 1.3336 |
| $O_{61}$ | 0.9295 | 0.0979 | 0.4412 | $O_{19}$ | 0.8627 | 0.9959 | 0.4309 | -0.0667 | -0.1020 | -0.0102 | 0.7941 |
| $O_{62}$ | 0.0979 | 0.9295 | 0.0588 | $O_{25}$ | 0.1409 | 0.9940 | 0.0535 | 0.0431 | 0.0646 | -0.0052 | 0.5059 |
| $O_{71}$ | 0.7239 | 0.5809 | 0.5651 | $O_{21}$ | 0.6607 | 0.6864 | 0.6130 | -0.0632 | 0.1055 | 0.0480 | 1.3182 |
| $O_{72}$ | 0.5809 | 0.7239 | 0.9349 | $O_{13}$ | 0.4705 | 0.6707 | 0.9023 | -0.1104 | -0.0532 | -0.0325 | 0.9161 |
| $O_{81}$ | 0.5996 | 0.0810 | 0.3914 | $O_{17}$ | 0.6331 | 0.1372 | 0.4309 | 0.0335 | 0.0562 | 0.0396 | 0.5847 |
| $O_{82}$ | 0.0810 | 0.5996 | 0.1086 | $O_{24}$ | 0.0059 | 0.6469 | 0.0535 | -0.0751 | 0.0473 | -0.0550 | 1.1172 |
| $O_{91}$ | 0.9277 | 0.3899 | 0.3821 | $O_{18}$ | 0.0040 | 0.3668 | 0.4309 | 0.0764 | -0.0230 | 0.0489 | 0.9573 |
| $O_{92}$ | 0.3899 | 0.9277 | 0.1179 | $O_{26}$ | 0.3530 | 0.8590 | 0.0535 | -0.0368 | -0.0687 | -0.0643 | 0.8309 |

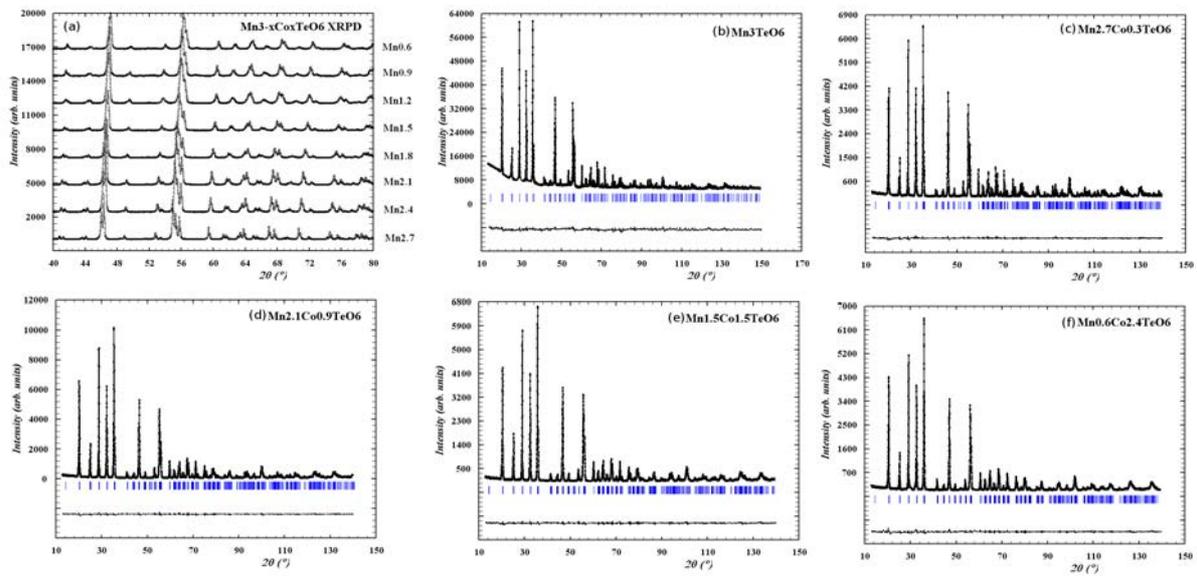

Figure 1

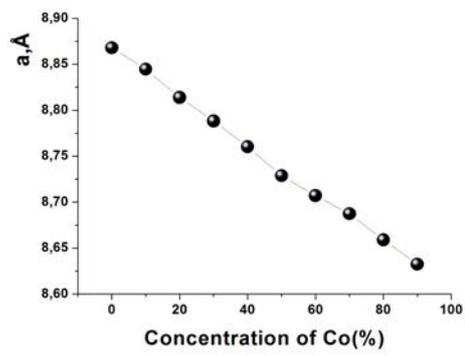
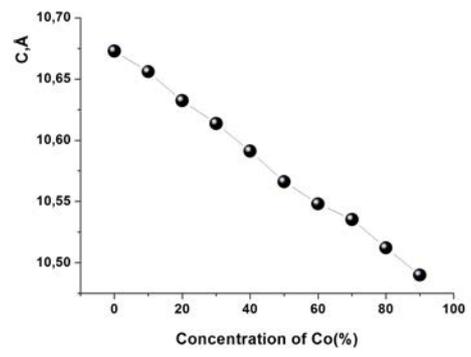
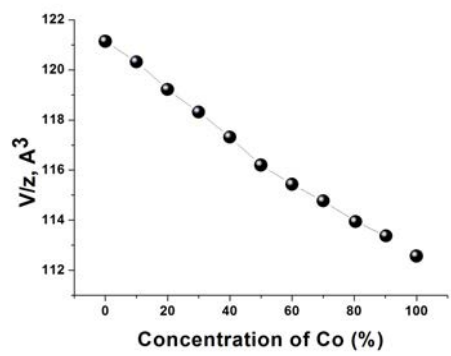

Figure 2

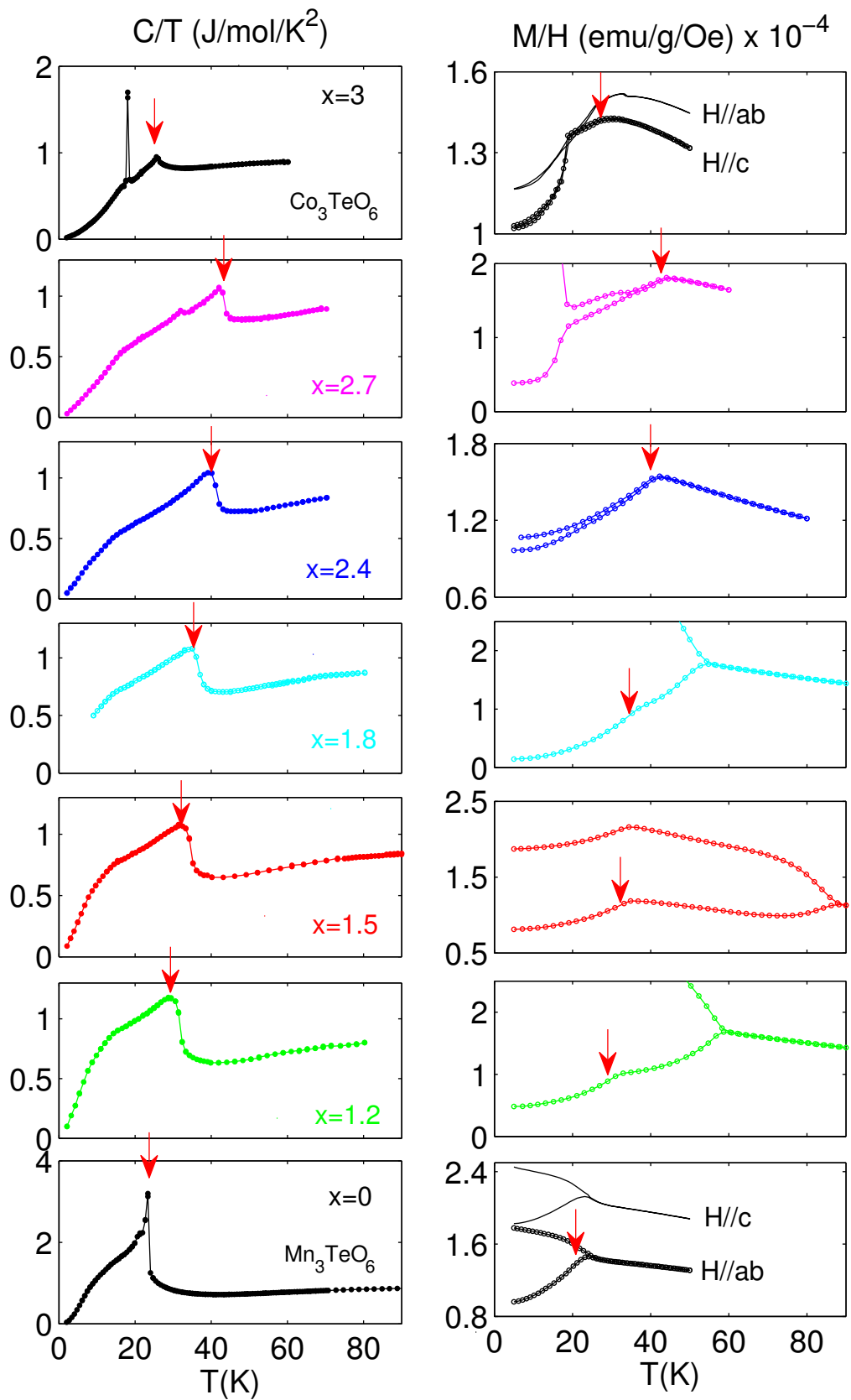

Figure 3

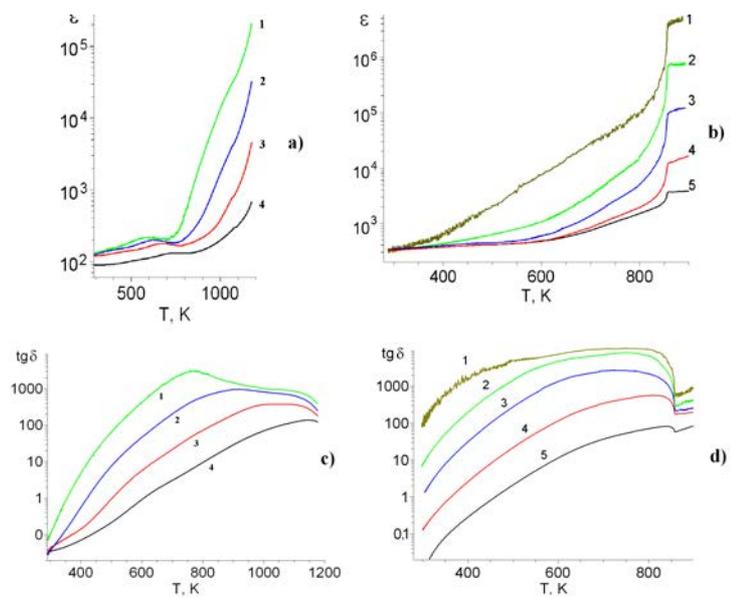

Figure 4

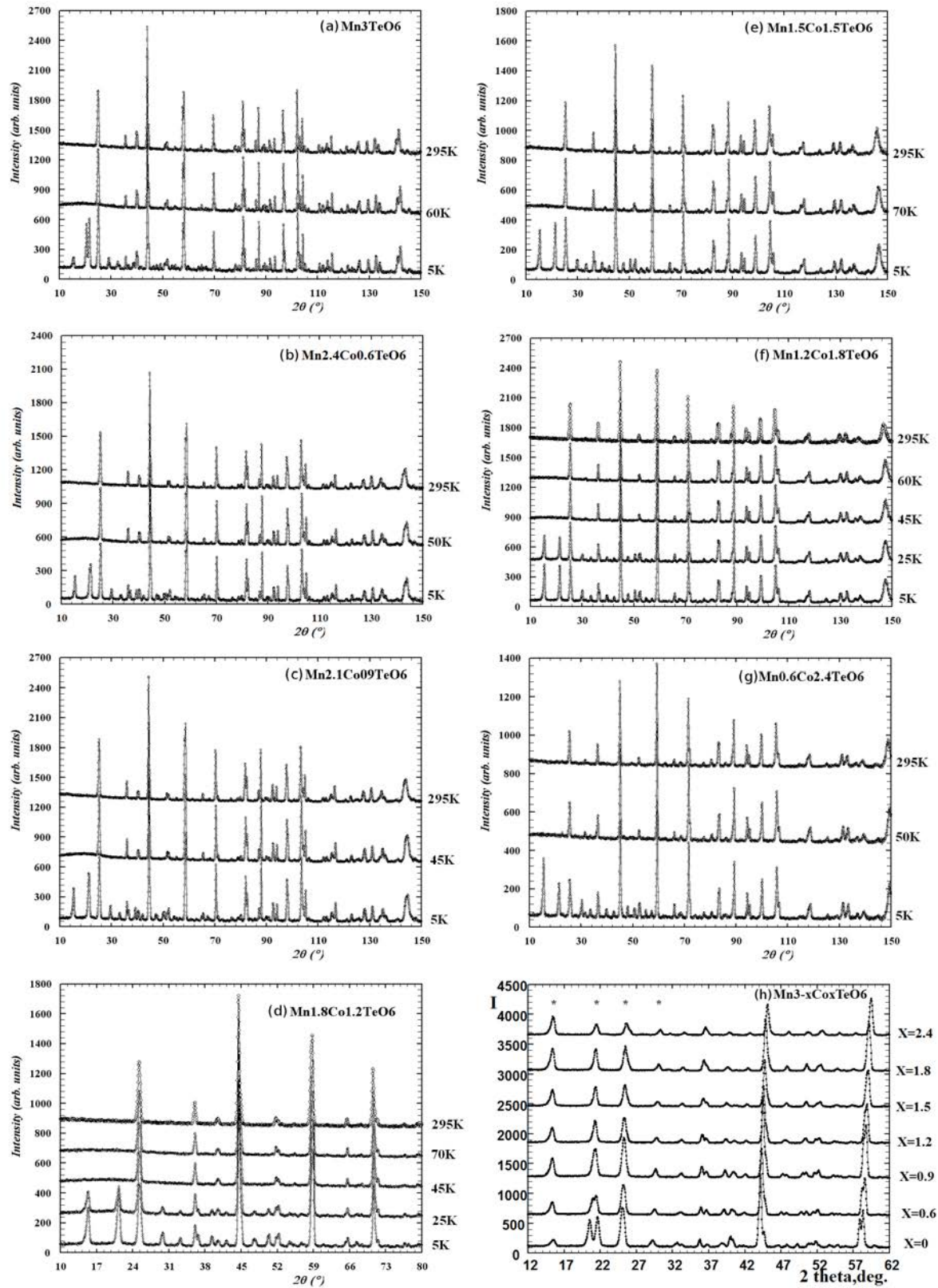

Figure 5

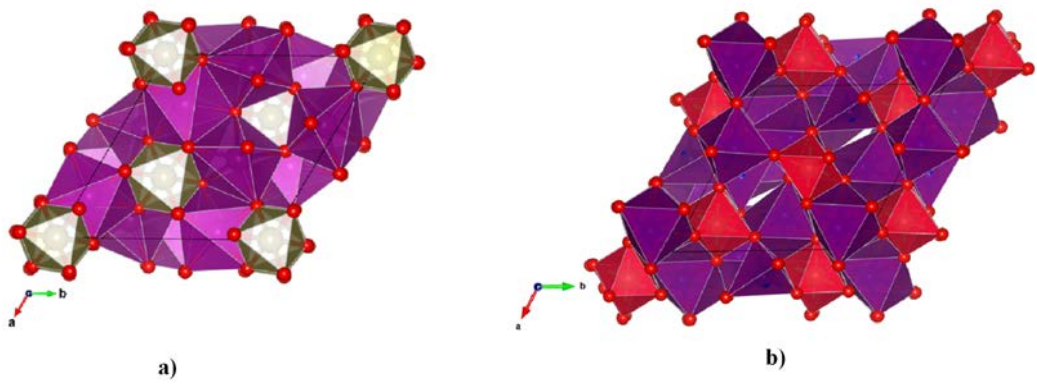

Figure 6

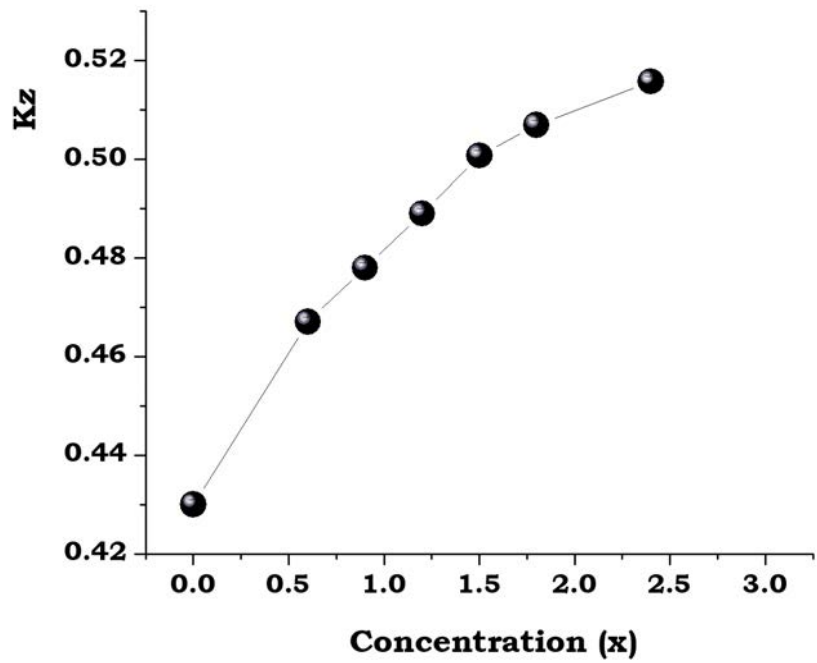

Figure 7

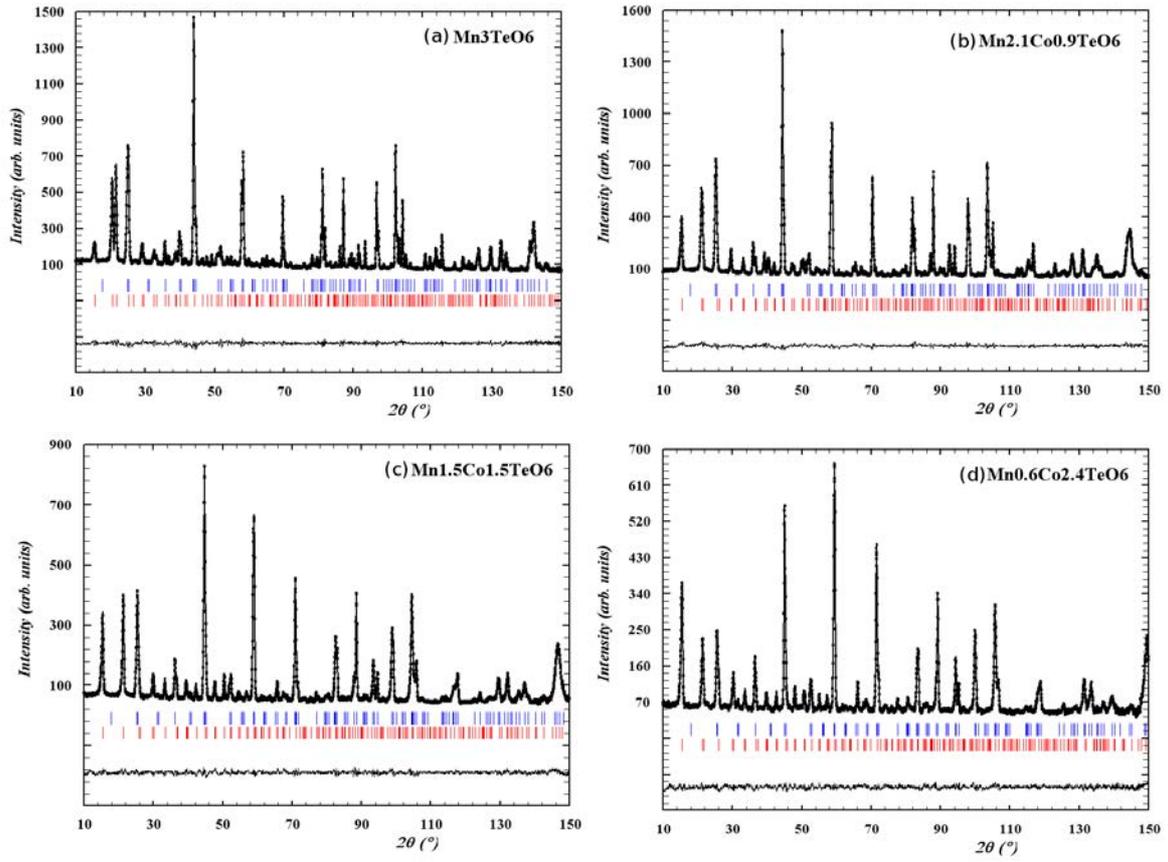

Figure 8

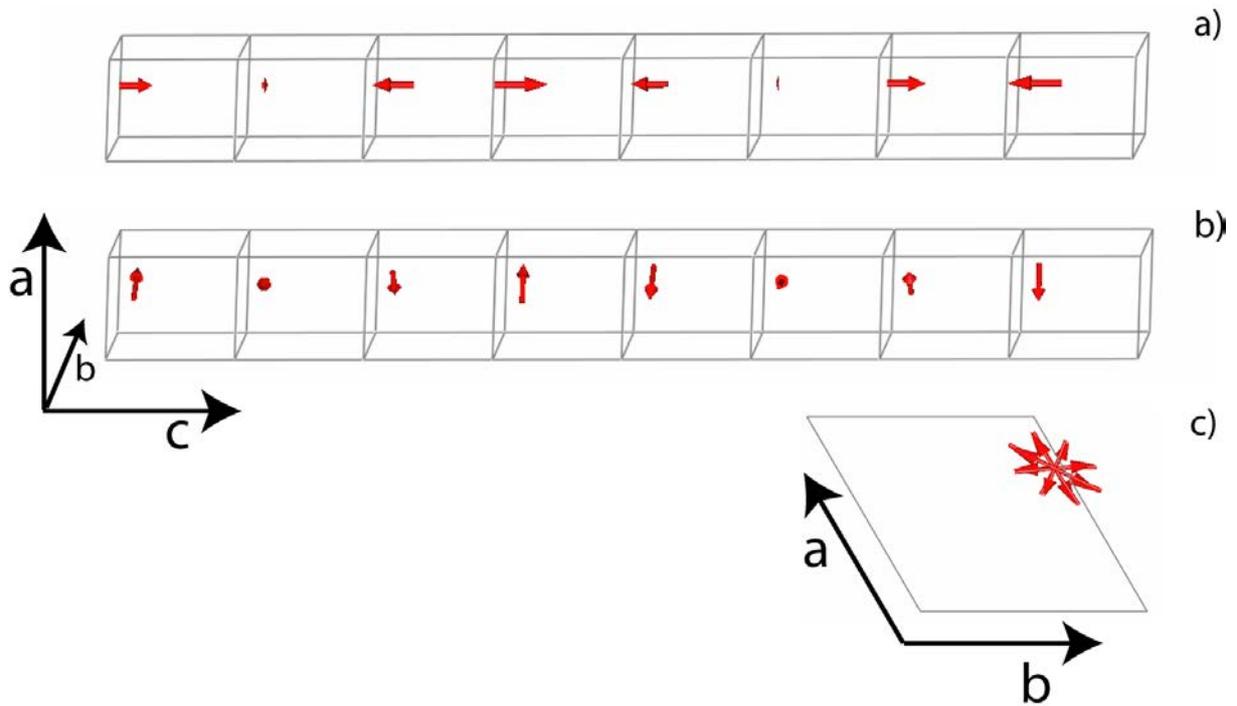

Figure 9

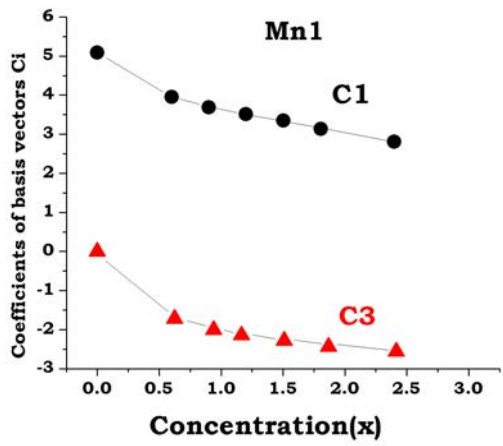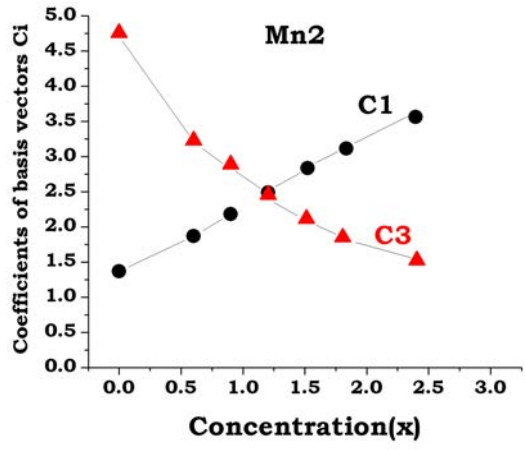

Figure 10

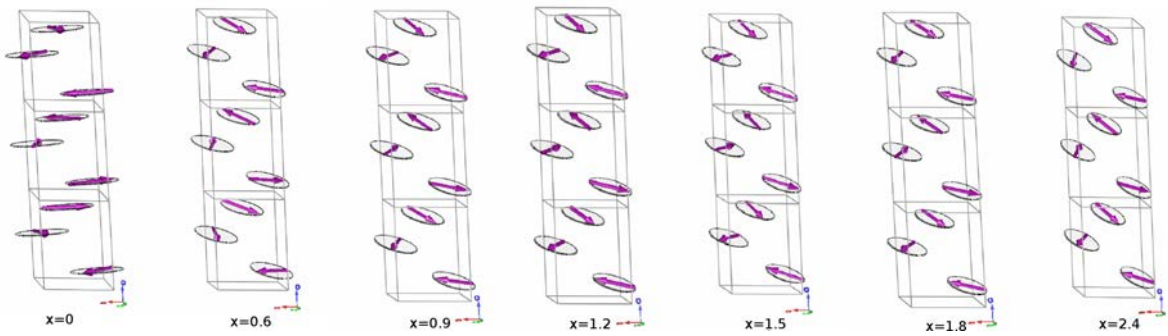

Figure 11

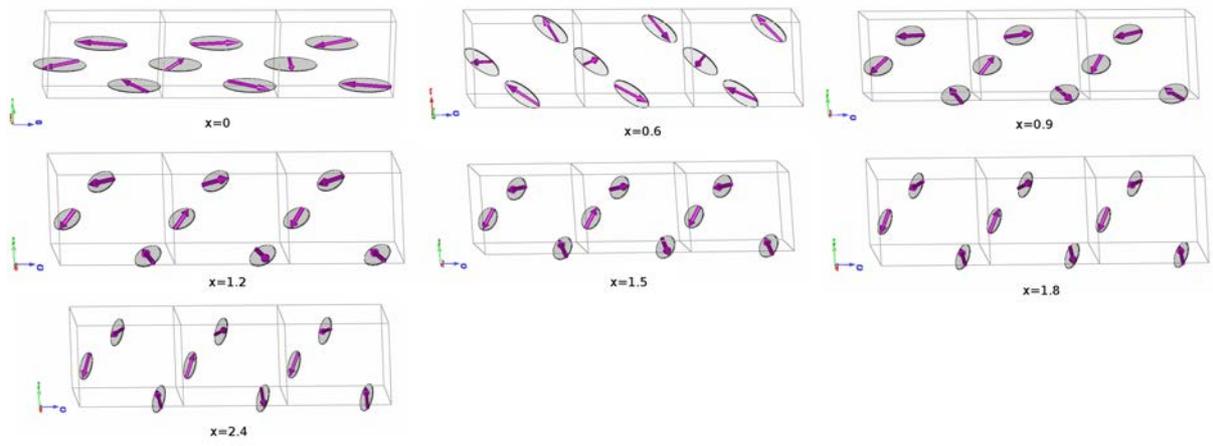

Figure 12

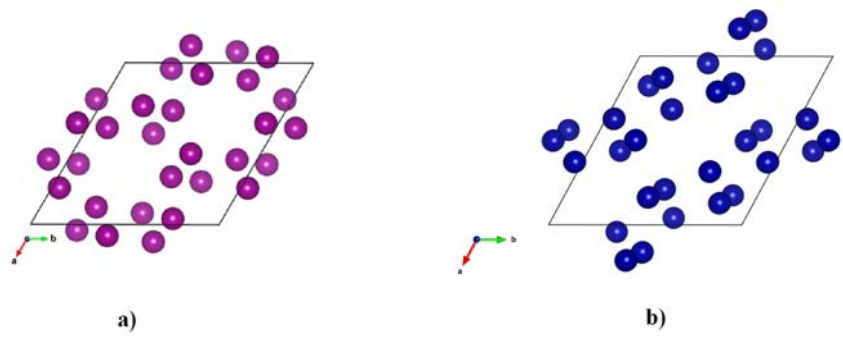

Figure 13